\documentclass[preprint,showpacs,english,aps]{revtex4-1}

\usepackage{babel,rotating,dcolumn}
\usepackage{colordvi,graphicx,color,amsbsy,amsmath,bm,amssymb}
\usepackage{comment}

\usepackage{graphicx}
\usepackage{epstopdf, epsfig}

\usepackage{babel,rotating,dcolumn}
\usepackage{colordvi,graphicx,color,amsbsy,amsmath,bm,amssymb}
\usepackage{comment}
\usepackage[normalem]{ulem}

\usepackage{natbib}

\usepackage{todonotes}

\usepackage{soul}

\DeclareSymbolFont{matha}{OML}{txmi}{m}{b}
\DeclareMathSymbol{\varv}{\mathord}{matha}{118}
\usepackage{mathrsfs}

\DeclareMathAlphabet\mathbfcal{OMS}{cmsy}{b}{n}

\usepackage{yhmath}
\usepackage{ucs}
\usepackage{amsfonts}

\usepackage{amsmath,esint}

\usepackage{nomencl}

\usepackage{bbm}

\newcommand{\thibault}[1]{\textcolor{black}{#1}}
\newcommand{\kai}[1]{\textcolor{black}{#1}}

\begin{document}


\title[]{\thibault{Neural network} models for preferential concentration of particles in two-dimensional turbulence}


\author{Thibault Maurel--Oujia$^{1}$}
\email{thibault.oujia@univ-amu.fr}
\author{Suhas S. Jain$^{2,3}$} 
\author{Keigo Matsuda$^{4}$} 
\author{Kai Schneider$^{1}$}
\author{Jacob R. West$^{5}$} 
\author{Kazuki Maeda$^{6}$} 
\affiliation{$^{1}$ Institut de Mathématiques de Marseille, Aix-Marseille Université, CNRS, Marseille, France}
\affiliation{$^{2}$ Center for Turbulence Research, Stanford University, Stanford, CA, USA}
\affiliation{$^{3}$ George W. Woodruff School of Mechanical Engineering, Georgia Institute of Technology, Atlanta, GA, USA}
\affiliation{$^{4}$ Japan Agency for Marine-Earth Science and Technology (JAMSTEC), Japan}
\affiliation{$^{5}$ Department of Mechanical Engineering, Stanford University, Stanford, CA, USA}
\affiliation{$^{6}$ School of Aeronautics and Astronautics, Purdue University, West Lafayette, IN, USA}
\date{\today}


\begin{abstract}
Cluster and void formations are key processes in the dynamics of particle-laden turbulence. 
In this work, we assess the performance of various neural network models for synthesizing preferential concentration fields of particles in turbulence. 
A database of direct numerical simulations of homogeneous isotropic two-dimensional turbulence with one-way coupled inertial point particles,  
is used to train the models using vorticity as the input to predict the particle number density fields.
We compare autoencoder, U--Net, generative adversarial network (GAN), and diffusion model approaches, 
and assess the statistical properties of the generated particle number density fields. 
We find that the GANs are superior in predicting clusters and voids, and therefore result in the best performance. 
Additionally, we explore a concept of ``supersampling", where neural networks can be used to predict full particle data using only the information of few particles, which yields \thibault{promising perspectives}
for reducing the computational cost of expensive DNS computations by avoiding the tracking of millions of particles. 
We also explore the inverse problem of synthesizing the enstrophy fields using the particle number density distribution as the input at different Stokes numbers. 
Hence, our study also indicates the potential use of neural networks to predict turbulent flow statistics using experimental measurements of inertial particles.

\end{abstract}


\maketitle

\section{Introduction}


The use of various tools and techniques from machine learning 
is becoming increasingly popular in the field of fluid mechanics 
\citep[e.g.,][]{duraisamy2019turbulence, brunton2020machine}. 
These tools are advancing everyday, 
and evaluating their potential and capabilities, 
especially of the deep neural networks, is critical to
improve and maximize
the use of data for modeling and analysis of various flows. 
In particular, for particle-laden flow, \citet{siddani2021machine} used a combination of convolutional neural networks and generative adversarial neural networks to recreate particle-resolved flow fields around a random distribution of particles, and for closure modeling for forces on particles in \citet{siddani2023point}. 
Machine learning was also used as described in \citet{faroughi2022meta} to predict the drag coefficient of spherical particles translating in viscoelastic fluids and in \citet{hwang2021machine,hwang2023deep} to model the forces and in \citet{hwang2022machine} to model collision on the nonspherical and irregular particles. 
For synthesizing flow data, a method using harmonic wavelet phase covariance has shown high-quality results in modeling geometric structures in turbulent flows \citep{zhang2021maximum} and inertial particles in turbulence \citep{brochard2022particle}. 

These and other previous studies use machine learning to develop models for the fine-scale dynamics of the flow and particles from the macroscopic scale of information, while the motivation of the present work is to use machine learning to inform on the mesoscale clustering of inertial particles in turbulence from flow field data without the need for information on individual particles' position and velocity.
Clusters of particles and void regions are characteristic features 
of particle-laden turbulent flows,
and their prediction and understanding are essential in various industrial applications. \thibault{This includes modeling radiative heat transfer in particle-laden solar receivers, impacting radar signal processing in clouds, and influencing the dispersion of pollutants in the atmosphere as well as droplets in combustion processes.} \kai{For a recent review we refer to \citet{brandt2022particle}.} 

\begin{figure} 
\centering
\includegraphics[width=0.85\linewidth]{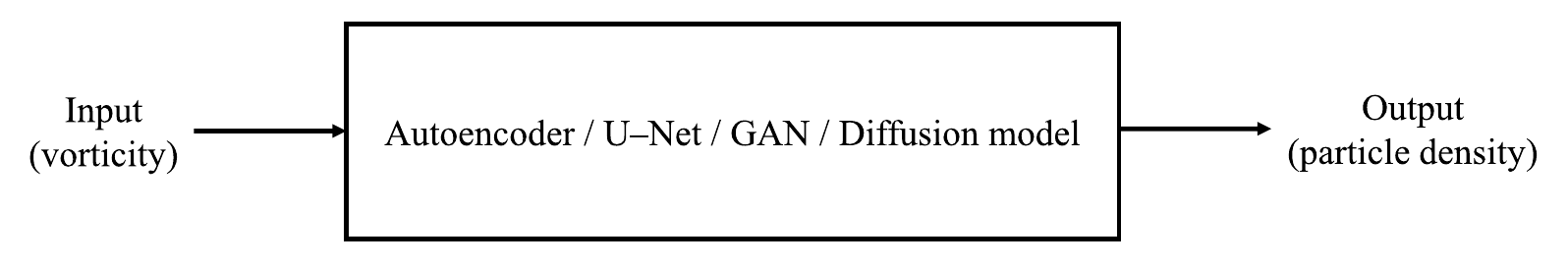}
\caption{A schematic illustrating the machine learning model\kai{s}. }
\label{fig:schematic}
\end{figure}

Measuring individual particles is typically impractical. In simulations,
it can be time-consuming to compute the trajectory of billions of particles.
Hence, it is imperative to develop an alternative low-cost technique to predict Eulerian description of particle distributions with varying 
parameters, e.g., increasing the number of particles or changing the particle inertia, without the need for tracking these large number of particles. 
As a first step, the prediction of particle distributions with the same parameters as the given data is important. 
The goal of this work is to develop and test machine learning tools for this task, which is illustrated in Figure \ref{fig:schematic}. 
To this end, four different neural networks, autoencoder, U--Net, generative adversarial network (GAN), and diffusion model, are trained using direct numerical simulation (DNS) data.
The aim is to synthesize one-way coupled particle number densities for different Stokes numbers from snapshots of enstrophy distributions of two-dimensional drift-wave turbulence, 
obtained by 
DNS \citep{kadoch2022lagrangian}. 
The results are then compared against the point-particle DNS data and statistically assessed.
In return for high-fidelity training data, which is expensive, the present results will contribute toward the development of efficient techniques that avoid expensive tracking of 
huge numbers
of particles in DNS computations. 
%

A complementary approach,  which is also considered here, is to invert the above procedure, i.e., particle number densities are used as input for generating different flow quantities, e.g., enstrophy, as output that is of some interest for experimentalists.
In fact, the enstrophy is directly related to the energy dissipation, which is a quantity of major interest. 

The remainder of the 
manuscript is organized as follows. Section~\ref{sec:methodology} presents the neural network methodology along with the details on the flow data set. Results are presented in Section~\ref{sec:results}, followed by the conclusion and perspectives for future work in Section~\ref{sec:conclusions}.

\section{Methodology}
\label{sec:methodology}

In this section, we provide details on the data set used, as well as a description of the neural network architecture and the training procedure.
We assess different neural networks in this work, and the motivation behind using different networks is to assess if more complex architectures yield improved results by generating fields that have better quality with respect to the DNS data. However, a trade-off is certainly necessary in terms of computational cost and accuracy.
The vorticity is used as the input to predict particle number density fields. A physical explanation behind using vorticity as the input quantity to generate the particle number density fields is the dynamics of inertial particles in turbulence. 
\thibault{
Inertial particles in turbulent flows exhibit a non-uniform spatial distribution, predominantly influenced by centrifugal forces. 
When the inertia of particles is small but finite, 
they tend to concentrate in regions of low vorticity magnitude and high strain rates, a phenomenon known as preferential concentration \citep{squires1990particle, squires1991preferential}. This effect arises because inertial particles in vortical regions are subjected to significant centrifugal forces, leading to their ejection. Conversely, in regions of straining with low vorticity, particles are more likely to accumulate due to reduced centrifugal expulsion. 
}
Other quantities related to \kai{inertial} particle dynamics, such as divergence or curl of the particle velocity, see e.g., \citet{maurel2023computing}, could be likewise predicted. But in this work, we choose to focus on density, which is a more fundamental quantity.


\subsection{Data set description}

\begin{figure} 
\centering
(a)\includegraphics[width=0.40\linewidth]{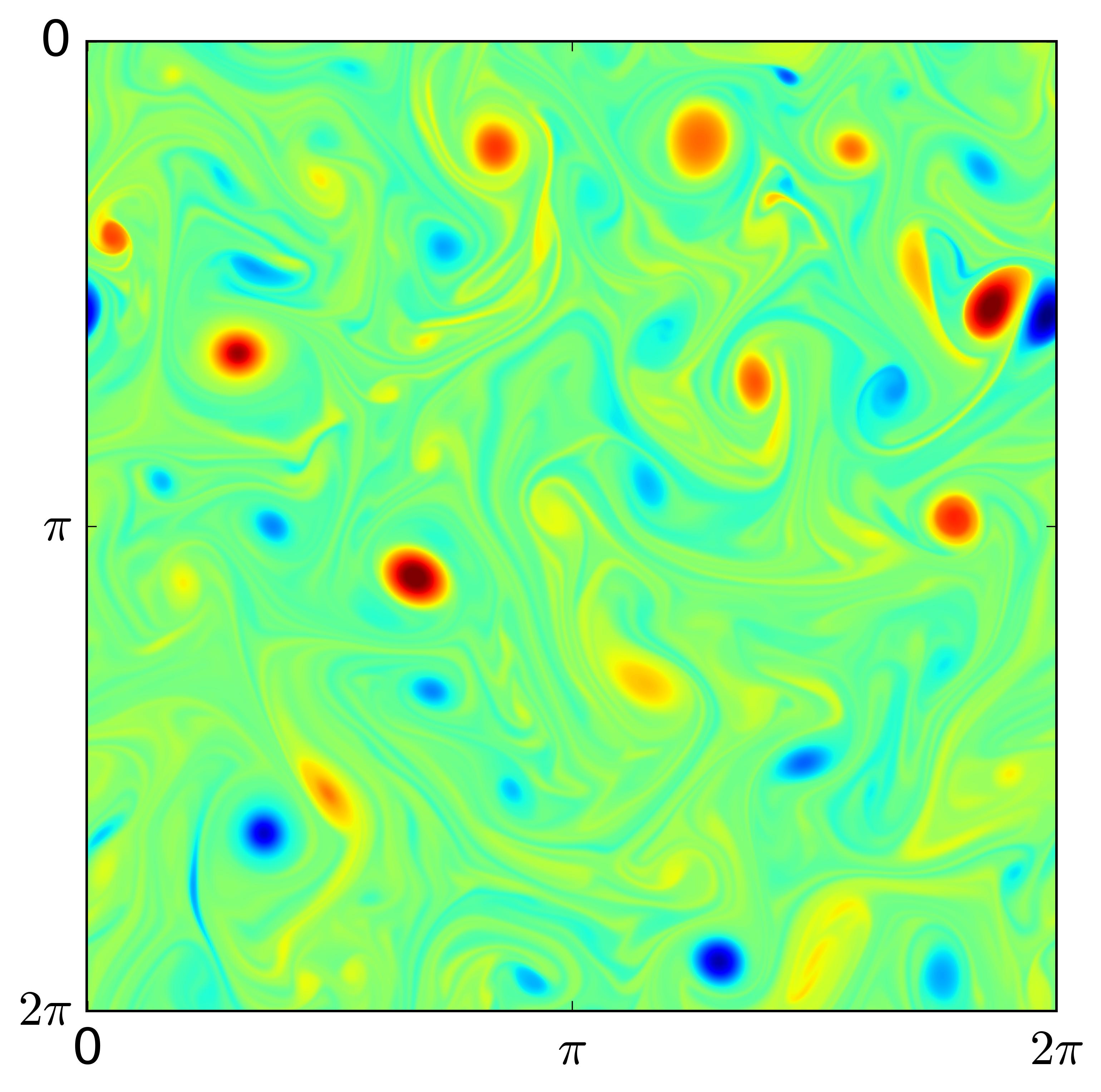}
(b)\includegraphics[width=0.40\linewidth]{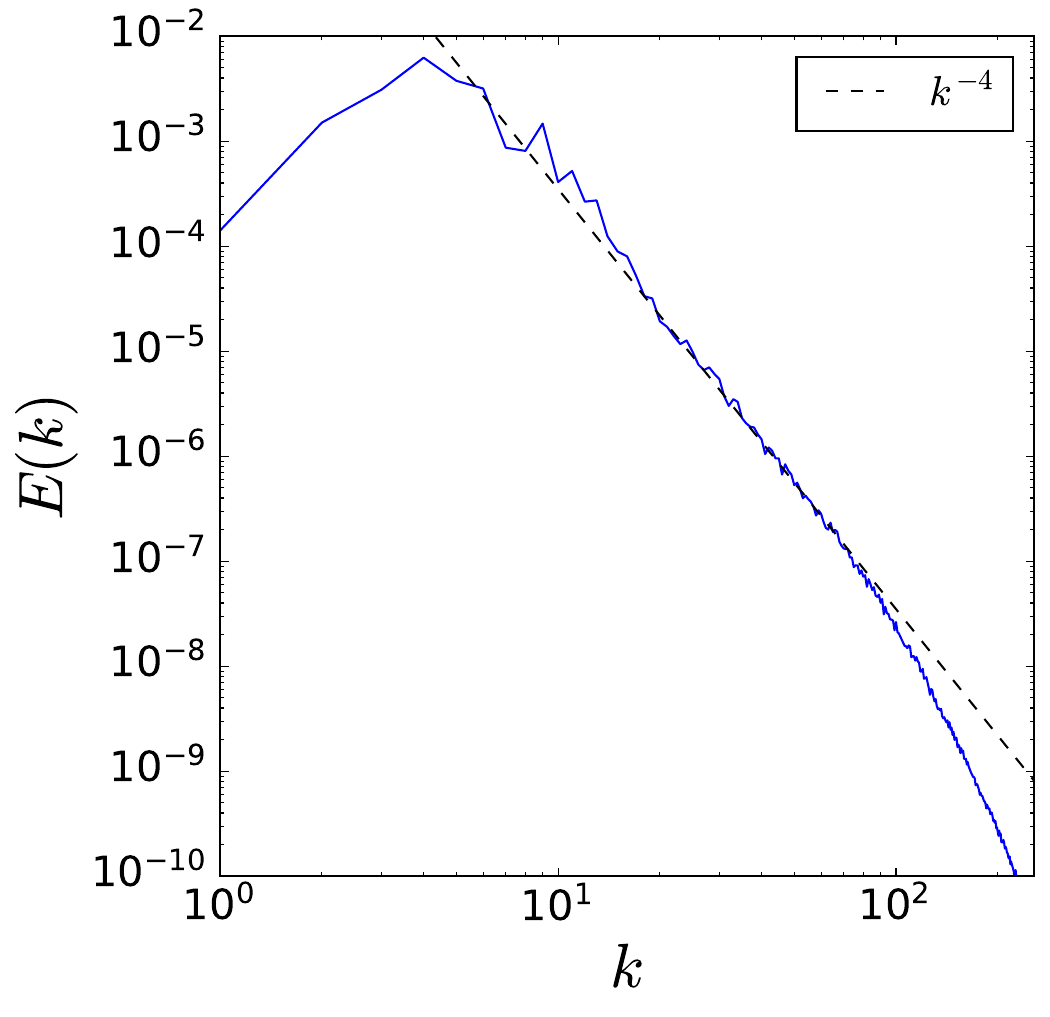}\\
\hspace{0.78cm}\includegraphics[width=0.39\linewidth]{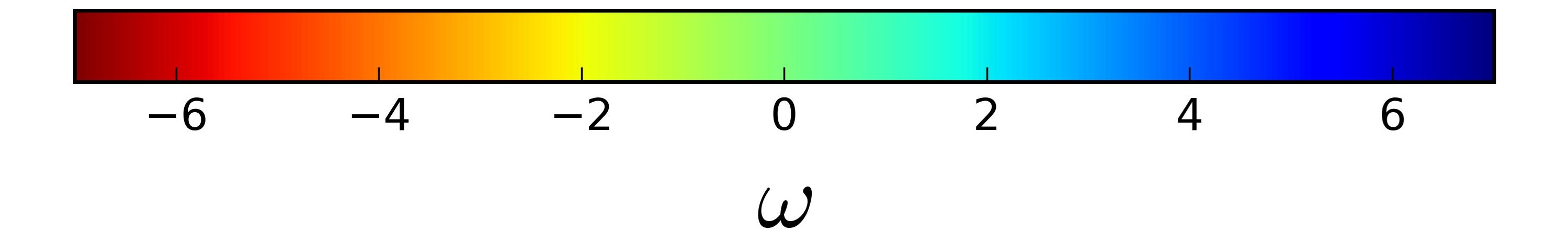}
\hspace{0.65cm}\includegraphics[width=0.39\linewidth]{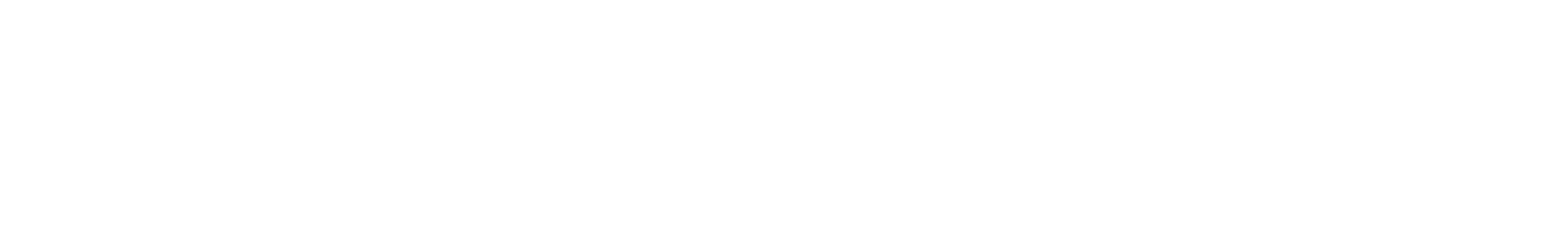}
\caption{Visualization of (a) the vorticity field of DNS data and (b) the corresponding energy spectrum in the statistically stationary regime.}
\label{fig:visualizations_vorticity_energy}
\end{figure}

The underlying data set consists of particle position and velocity data generated by DNS of particle-laden drift-wave turbulence in the hydrodynamic regime, which is \kai{similar} to 2D Navier-Stokes turbulence, as detailed in \citet{kadoch2022lagrangian}. 
The governing equations are solved in a $2 \pi$-periodic 
square with a Fourier pseudo-spectral scheme.
%
The flow reaches a statistically stationary regime, which is close to 2D homogeneous isotropic turbulence with large scale forcing and exhibits a $k^{-4}$ turbulent kinetic energy spectrum as shown in Figure~\ref{fig:visualizations_vorticity_energy}. 
\thibault{\kai{The} observation of a $k^{-4}$ turbulent kinetic energy spectrum aligns with findings in the literature, notably by \citet{legras1988high}, where such spectral 
\kai{slopes where reported for 2D Navier--Stokes turbulence.}}
%
Uniformly distributed discrete particles are then injected into the fully developed flow and are tracked in the one--way coupled Lagrangian framework. 
Maxey's model \citep{Maxe87} for inertial heavy point particles with Stokes drag is used, and the inertial dynamics is represented by the Stokes number, $St = \tau_p / \tau_f $, where $\tau_p$ is the particle relaxation time and $\tau_f$ the \thibault{eddy turn over time}, \kai{defined by $\tau_f = \sqrt{\nu/ \epsilon}$ where $\nu$ is the kinematic viscosity and $\epsilon$ the dissipation rate.}
\thibault{
The position of the particle, denoted as ${\bm x}_p$, and its velocity, represented by ${\bm v}_p$, evolve according to the following equations:
\begin{equation}
    d_t {\bm x}_p = {\bm v}_p, \quad 
    d_t {\bm v}_p = - \frac{{\bm v}_p-{\bm u}_p}{\tau_p}
\end{equation}
where ${\bm u}_p$ refers to the velocity of the fluid at the location of the particle ${\bm x}_p$.
}
%

DNS with $N_g^2 = 1024^2$ grid points is performed for a Reynolds number, $Re_\lambda =  679$. For details we refer 
to \citet{kadoch2022lagrangian}. 
The number of particles $N_p$ is $10^6$ and the considered five Stokes numbers are $St =$ 0.05, 0.1, 0.25, 0.5, and 1, each corresponding to a different data set.
Particles with different Stokes numbers were tracked in an identical turbulent flow for about 13.72 eddy turn-over times.

We provide as input the vorticity fields and aim to predict the particle number density for various Stokes numbers. 
To facilitate faster convergence of the neural network by mitigating significant variations in output values, we perform clipping of density values exceeding 100, \thibault{then divide them by 100 to ensure that the values are normalized between 0 and 1. This clipping} results in the removal of just under 1\% of the total particle mass.
%
The possibility of performing super-sampling, that is, providing as input, not only the vorticity, but also the particle number density for an extremely low number of particles, is likewise explored.
We are also interested in doing the inverse, i.e., predicting the enstrophy from the particle number density for different Stokes numbers.
We have a total of 100 saved snapshots, and use the first 80 snapshots for the training and the last 10 for the test set. We leave a gap of 10 snapshots between the training and test sets to ensure sufficient variability and decorrelation between the two. 
The particle number density is calculated with a resolution of $512^2$.
Similarly, we use the vorticity 
field, to which average pooling has been applied, with a resolution of $512^2$ as input.

Moreover, we artificially increase the size of the training data eight times by applying different operations of transposition and rotation to the original data. 
This yields in total a data training set of $80 \times 8 = 640$ images and a test set of $10 \times 8 = 80$ images, for each Stokes number.

\subsection{Description of the neural network architecture and training}

\begin{figure} 
\centering
\includegraphics[width=\linewidth]{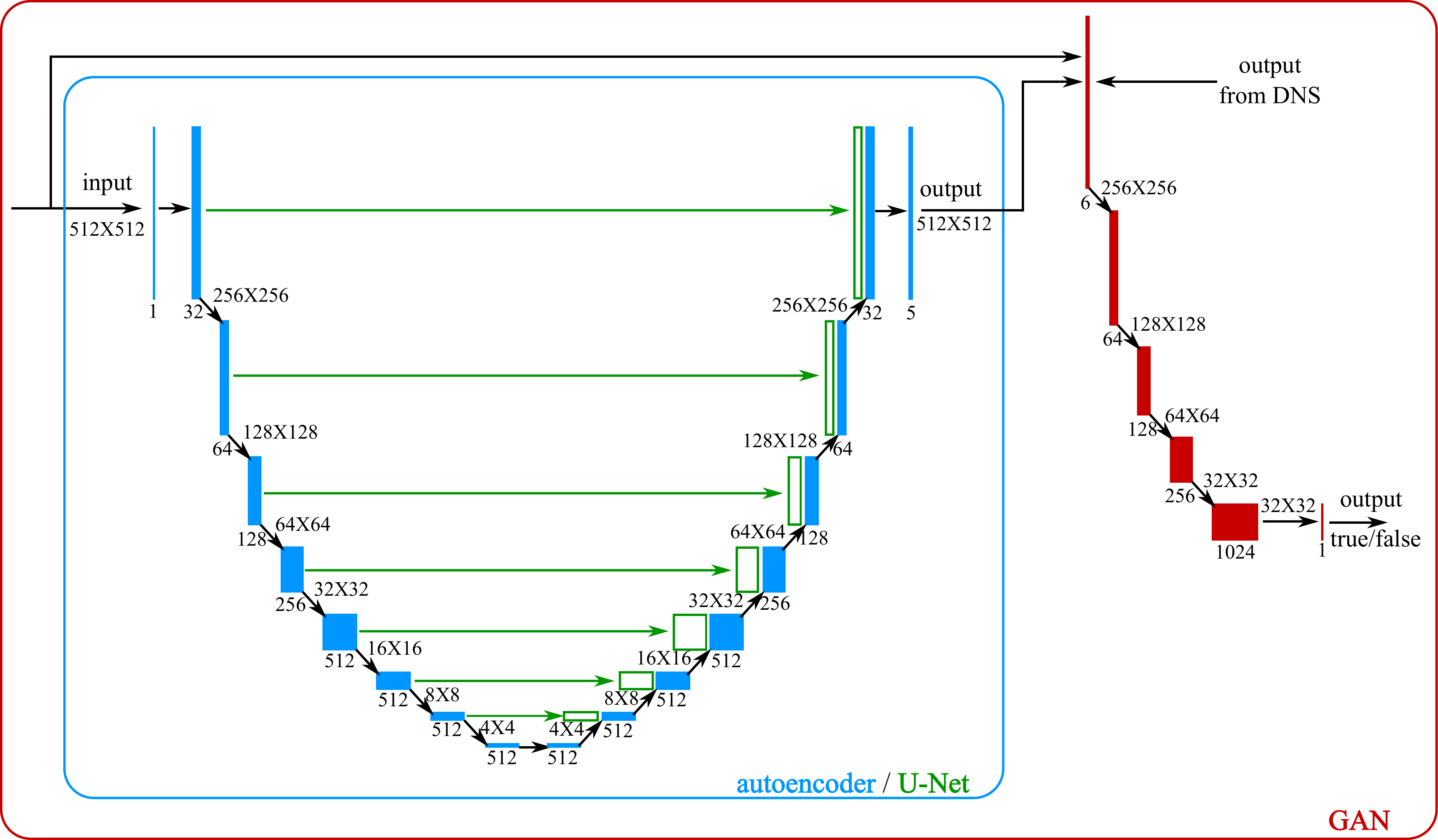}
\caption{Illustration of different neural network architectures: autoencoder (blue), U--Net (blue and green) and GAN (blue, green and red). 
The U--Net architecture is also utilized for the diffusion model.
}
\label{fig:CNN}
\end{figure}

Figure~\ref{fig:CNN} shows the different architectures we have tested to synthesize the number density distribution of the inertial particles for different Stokes numbers. We compare four different neural networks: autoencoder, U--Net, GAN, and diffusion model, each of which is a modification of the standard architecture, adapted here for the current application.
An autoencoder \citep{rumelhart1986learning} is an artificial neural network composed of two neural networks, an encoder and a decoder. The encoder compresses/reduces the size of the data, while the decoder reconstructs the desired output using the compressed data.
The autoencoder is the baseline network for this study and is represented by the blocks and arrows in blue in Figure \ref{fig:CNN}.
The loss function used here is the binary cross-entropy between the exact and predicted values. 
A U--Net \citep{ronneberger2015u} is an autoencoder with additional skip connections between the layers of the encoder and the decoder, which have the same shape. 
Skip connections bypass some of the layers in the neural network to allow a connection between two distant layers, contrary to the standard connection, which are only connected to the next layer.
The U--Net is represented in Figure \ref{fig:CNN} as the blue part (autoencoder), to which we have added skip connections in green.
We use the same loss function (binary cross-entropy) as in the autoencoder.
A GAN \citep{goodfellow2020generative} 
is an artificial neural network composed of two neural networks, a generator and a discriminator. The generator learns to generate new data set, and the discriminator classifies if the input data is real or generated. The goal of the generator is to generate data that will be classified as real by the discriminator.
We use a U--Net as the generator, as is done in \citet{isola2017image}. We give as input to the discriminator the vorticity and the synthetic or exact particle number density.
For the discriminator, we use the binary cross-entropy between the correct and the exact classifications (generated or real image).
For the generator, we use the binary cross-entropy between the exact and the predicted values to which we add the loss of the discriminator in case we try to fool it.
The GAN is represented in the figure as the previous architecture to which we have added a discriminator in red.
Unlike previous architectures, the diffusion model \citep{ho2020denoising} operates iteratively through a series of denoising steps to incrementally refine the output. At each iteration, a noisy version of the particle number density is fed into the model, which is trained to predict the noise that has been added. The predicted noise is then subtracted from the noisy image, effectively achieving a slight denoising effect. The denoising loss is computed using an L1-norm between the denoised image and the ground truth. This process is repeated for 500 denoising steps to achieve a finely denoised output. We employ a U--Net as the denoising model to execute these steps. The iterative process enables the model to learn complex patterns in the data, providing a more accurate and reliable prediction of the particle number density distribution for various Stokes numbers. 
To condition the model, we also provide the noise-free vorticity 
field as input.

The down-sampling convolution blocks are composed of a 2D convolution with a $3\times3$ filter, followed by batch normalization and a leaky rectified linear unit (LeakyReLU) activation function.
To respect the spatial periodicity intrinsic to our problem, periodic padding is applied, thereby maintaining the data resolution.
The sequence is repeated twice in each block, with the inclusion of a max-pooling operation at the end of the convolution block to coarsen the spatial dimensions.
Conversely, the up-sampling convolutional blocks adopt a similar sequence but utilize upsampling operations to expand the spatial dimensions, effectively reversing the encoding process undertaken during down-sampling.

We use the Adam optimizer for training the networks \citep{kingma2014adam}. 
Training was executed on an Nvidia Tesla V100 32GB GPU. 
The number of epochs was chosen to ensure the convergence of the different neural networks tested.
For the autoencoder and the U--Net, we use 1000 epochs, which was more than sufficient for convergence. 
In the case of the GAN, due to the model's inherent instability, we selected an epoch with the lower loss value from among all the saved steps throughout the training process. 
For the diffusion model, we choose to train for 400 epochs, beyond which we observed signs of overfitting.
%

\section{Results}
\label{sec:results}

In this section, we present results for different neural networks considered in this study. We first analyze the networks using visualizations in physical space and then assess them statistically, i.e., we present probability distribution functions (PDFs) and particle number density Fourier spectra and compare them with the DNS data.
The presented results were obtained from the validation data set disjoint from the training data set, i.e., 
from different snapshots. 

 \subsection{Prediction of preferential concentration}

 \subsubsection{Some visualizations}

\begin{figure} 
\centering
(a)\includegraphics[width=0.40\linewidth]{FIGURES/vorticity_pos_neg_Exact.jpeg}
(b)\includegraphics[width=0.40\linewidth]{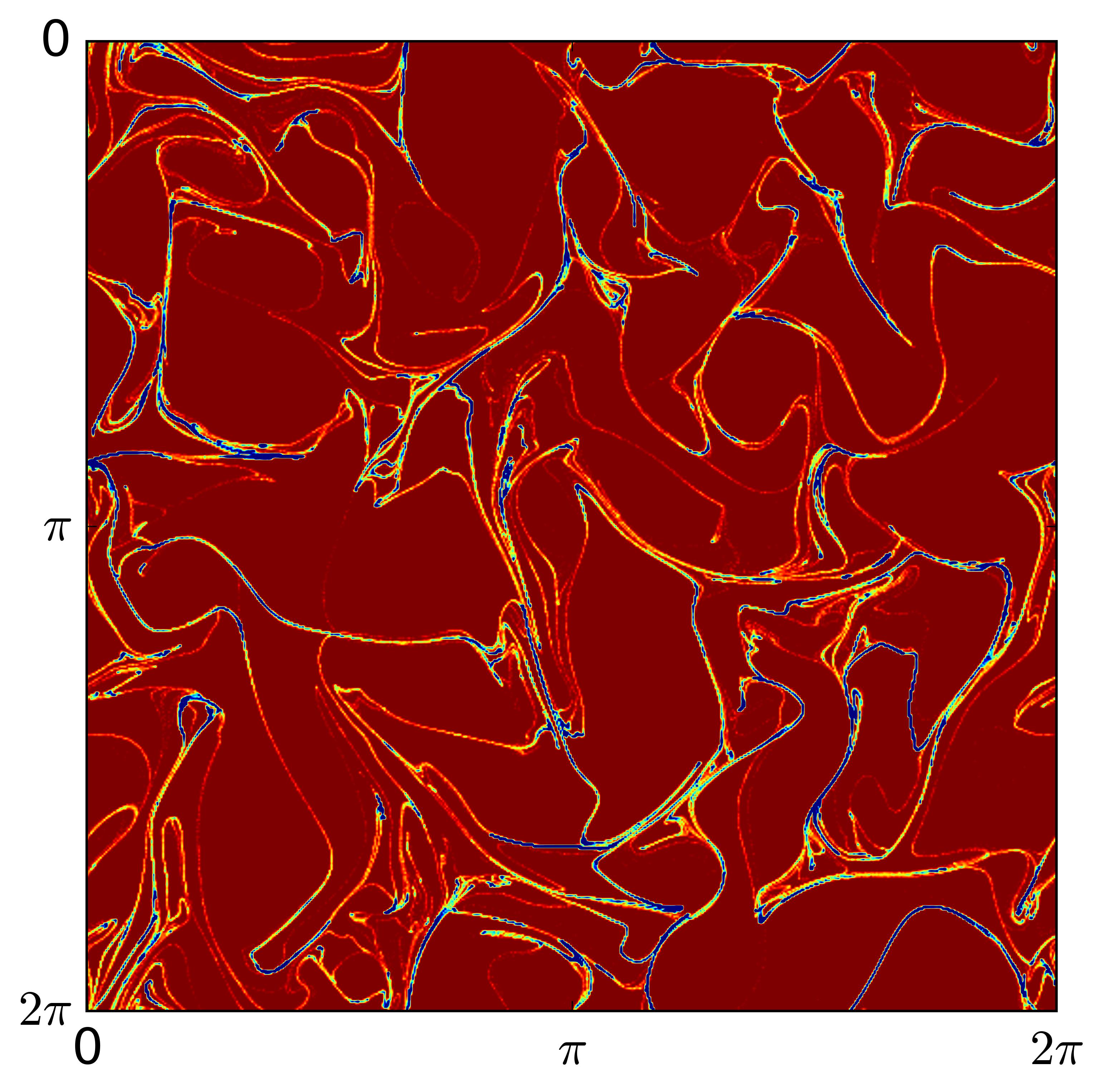}\\
\hspace{0.78cm}\includegraphics[width=0.39\linewidth]{FIGURES/colorbar_vorticity_pos_neg_Exact.jpeg}
\hspace{0.65cm}\includegraphics[width=0.39\linewidth]{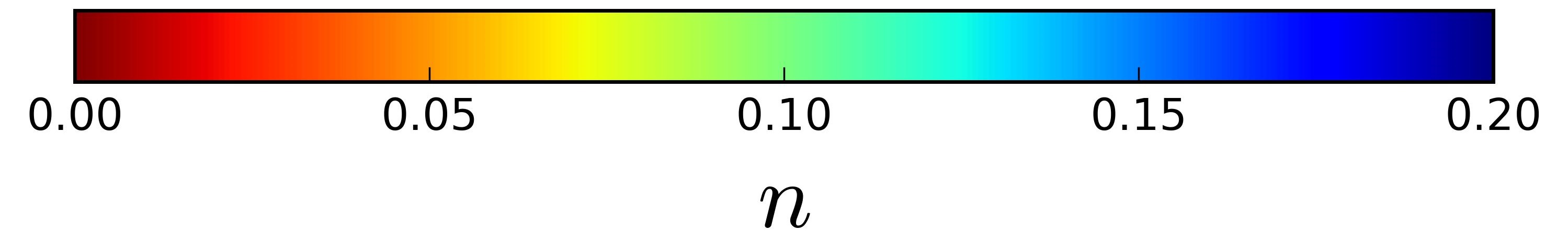}\\
(c)\includegraphics[width=0.40\linewidth]{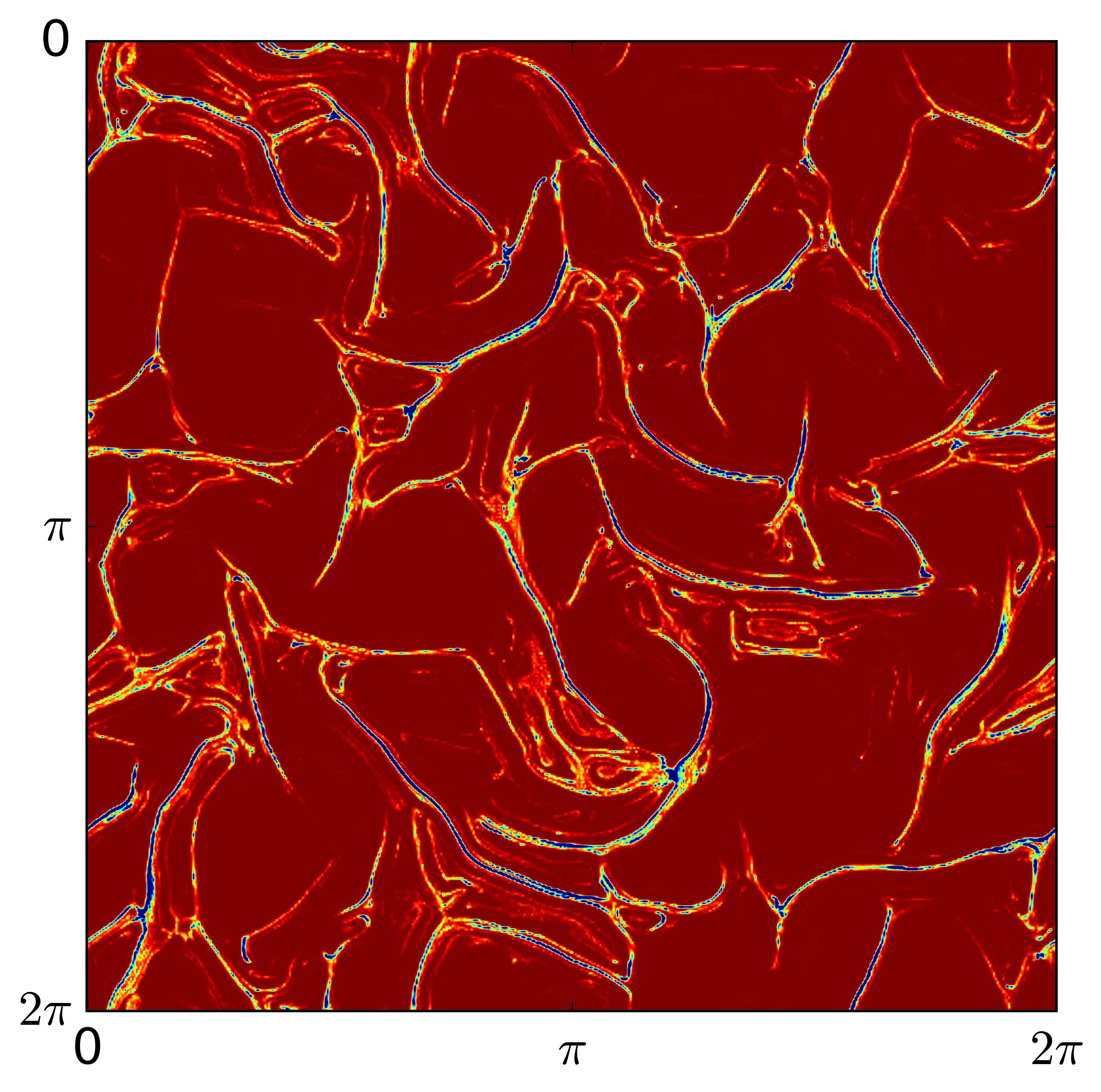}
(d)\includegraphics[width=0.40\linewidth]{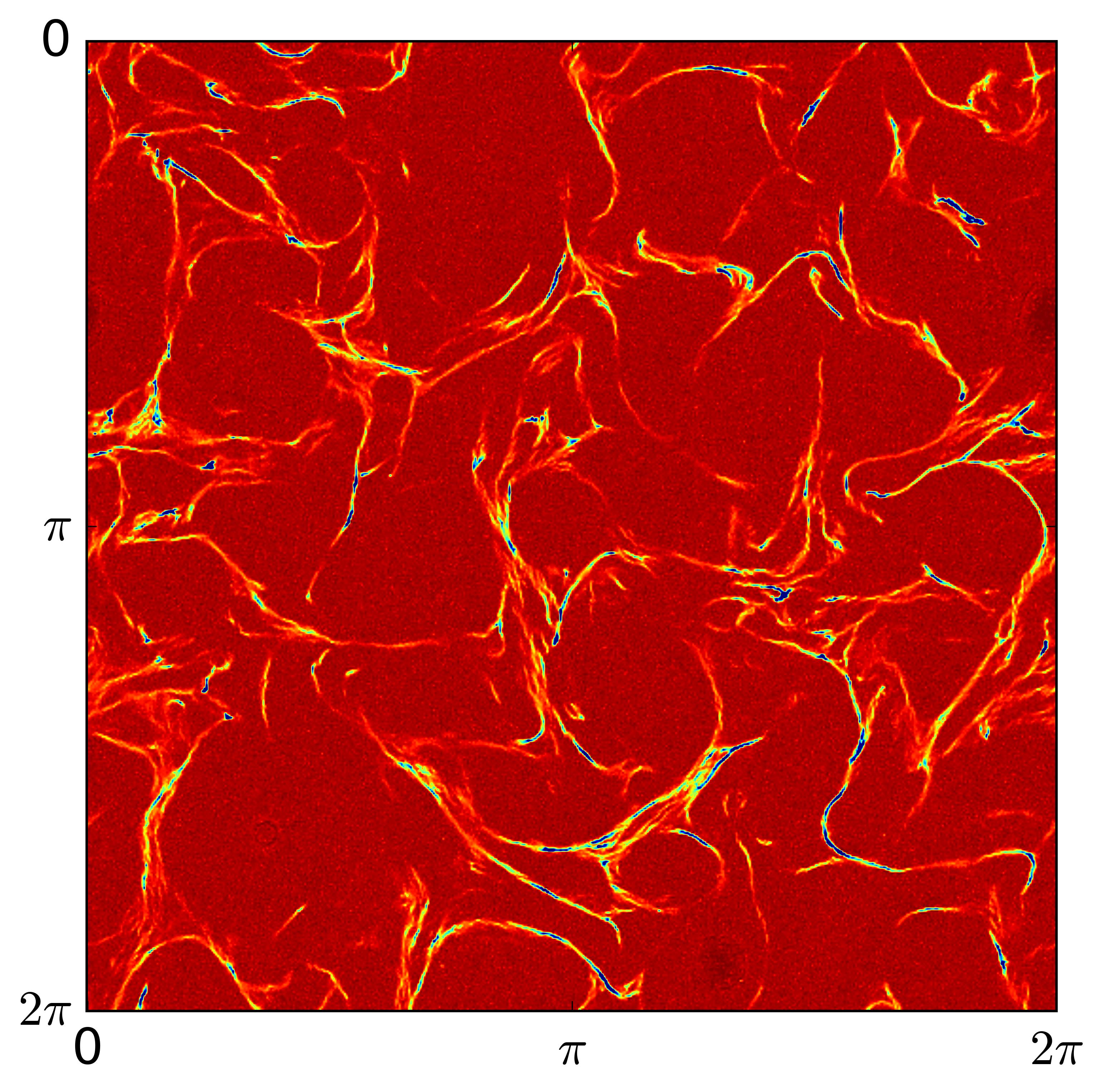}\\
(e)\includegraphics[width=0.40\linewidth]{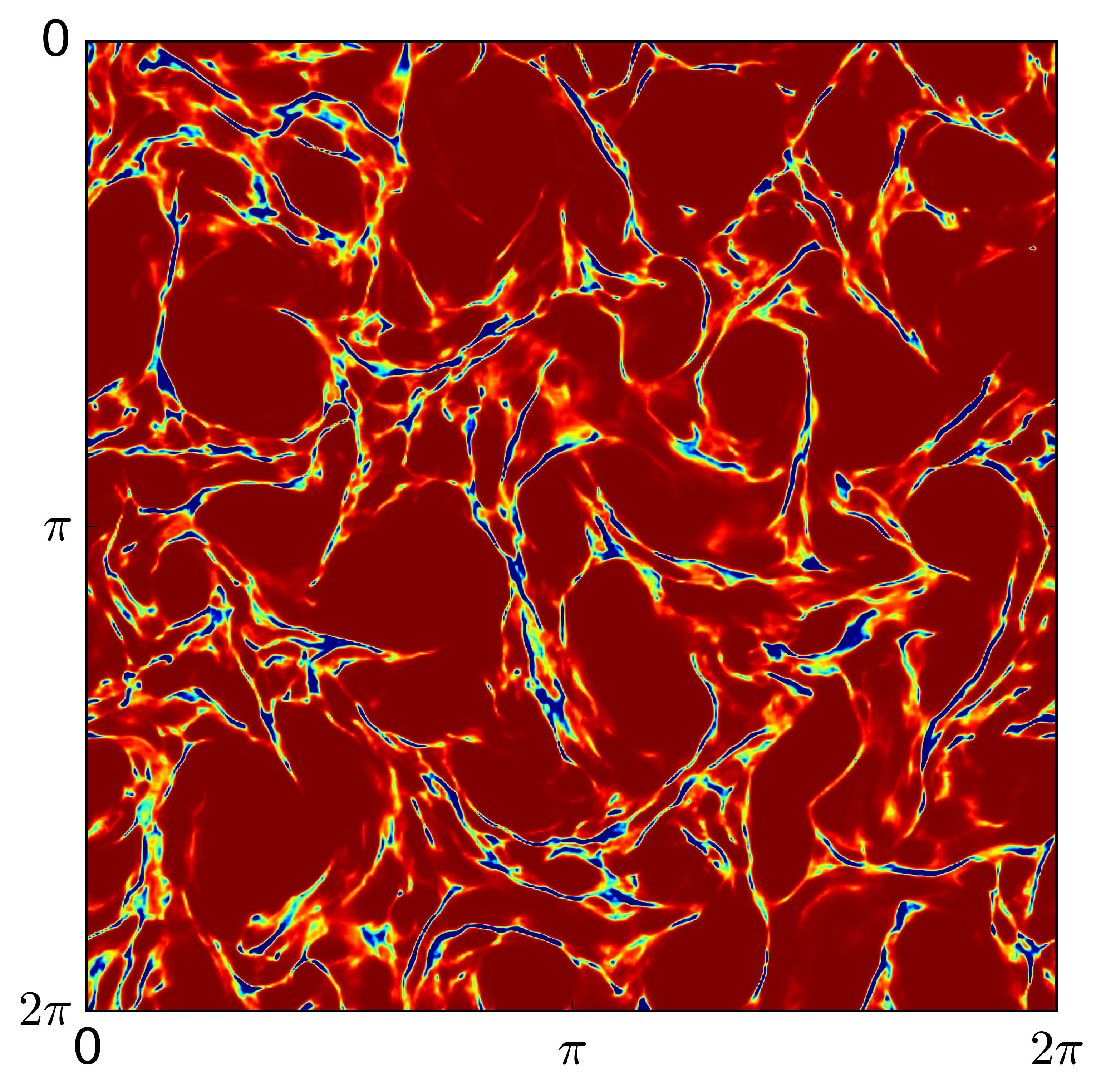}
(f)\includegraphics[width=0.40\linewidth]{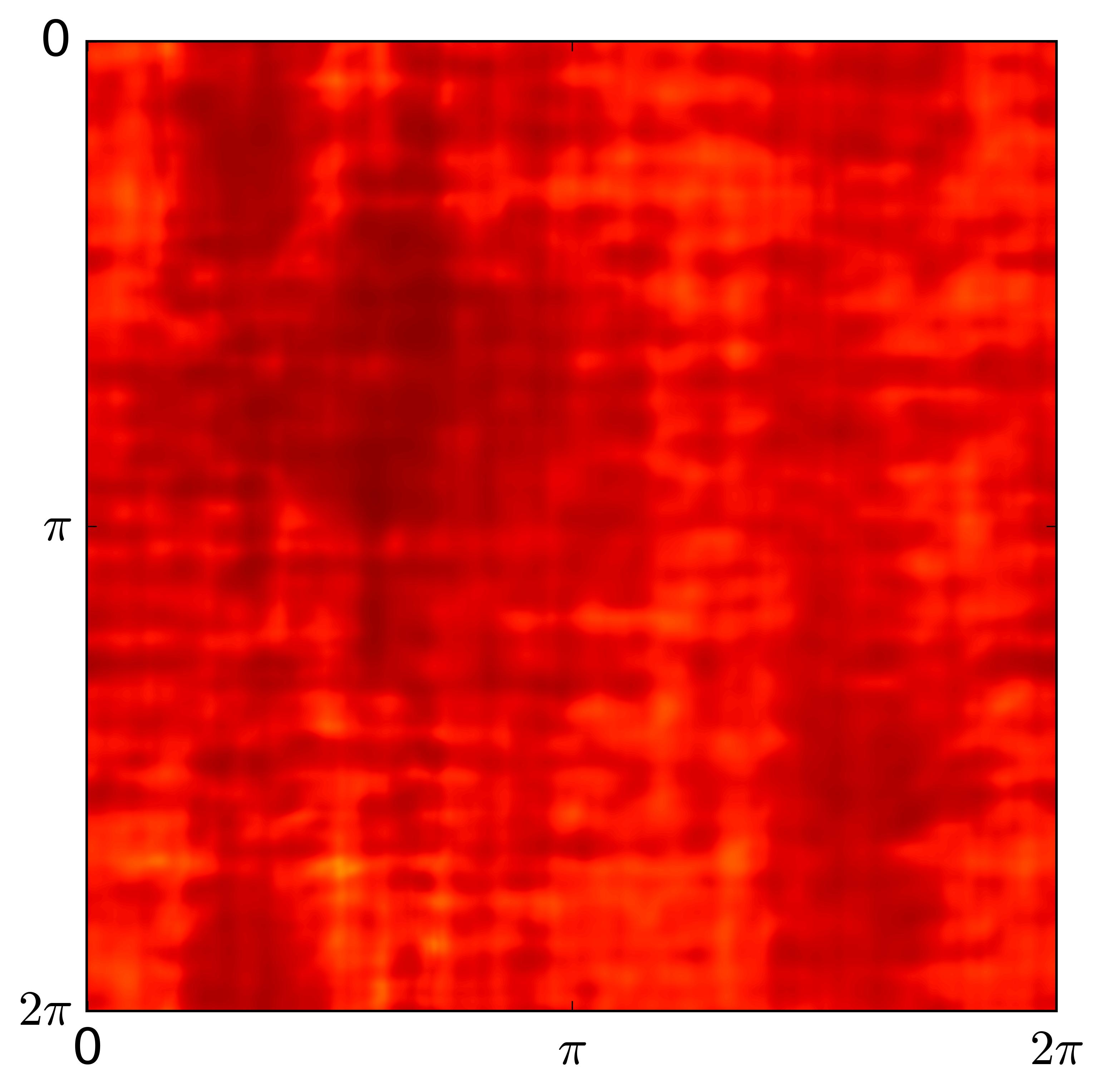}
\caption{(a) Input vorticity and (b) ground truth particle number density for $St = 1$ corresponding to DNS data. Predicted particle number density using (c) GAN, (d) diffusion model, (e) U--Net and (f) autoencoder.
}
\label{fig:visualizations_predictied_density}
\end{figure}


Figure \ref{fig:visualizations_predictied_density} shows (a) the input vorticity
field and (b) the ground truth particle number density for $St = 1$ as obtained from DNS, along with the predicted particle densities for $St = 1$ using (c) GAN, (d) diffusion model, (e) U--Net, and (f) autoencoder.
In the vorticity field, we can observe localized zones of high rotational velocity, contrasting sharply with surrounding areas of lower vorticity.
%
%
From the particle number density field, we can see that the particles are clustered in the regions of low vorticity magnitude
and avoid regions of high vorticity magnitude due to the centrifugal force, \thibault{similar to what is observed in \citet{pandey2019clustering}.}
The prediction of the particle number density field using the GAN shows similar structures in certain regions, while also displaying some dissimilarities. Overall, a lower density of particles is observed in regions of high vorticity, as expected. However, the model also predicts the presence of particles in some areas where none were observed in the DNS field.
The predicted particle number density fields also show the presence of particle clusters with thin filaments and sharp transitions from void regions to cluster regions. This is a sign that the GAN is able to generate fine-scale features accurately. 
This ability of GANs to predict fine-scale features has already been observed in \citet{ledig2017photo} in the context of image processing. 
In the diffusion model prediction, we observe a substantial amount of background noise, indicating that the model has not completely succeeded in filtering out the intrinsic noise from the training process. However, there are instances where the filamentary structures are accurately positioned, akin to the results obtained with the GAN. In other cases, the alignment is less precise. Despite this, fine filaments are observed which indicates the diffusion model's ability to generate rapid density variations, which are essential for the formation of void or cluster regions.
For the U--Net prediction, the void regions are generally located where they should be, specifically in areas of high vorticity. However, we do not observe the presence of fine filaments, as seen in the predictions from the GAN and the diffusion model.
In the autoencoder prediction, we notice a grid-like artifact; the particle number density appears diffused and blurry. The regions devoid of particles are indistinct from particle clusters. We conjecture that this could be due to the loss of information at various scales during the data compression stage.

\subsubsection{Comparison of statistics}

\begin{figure} 
\centering
(a)\includegraphics[width=0.45\linewidth]{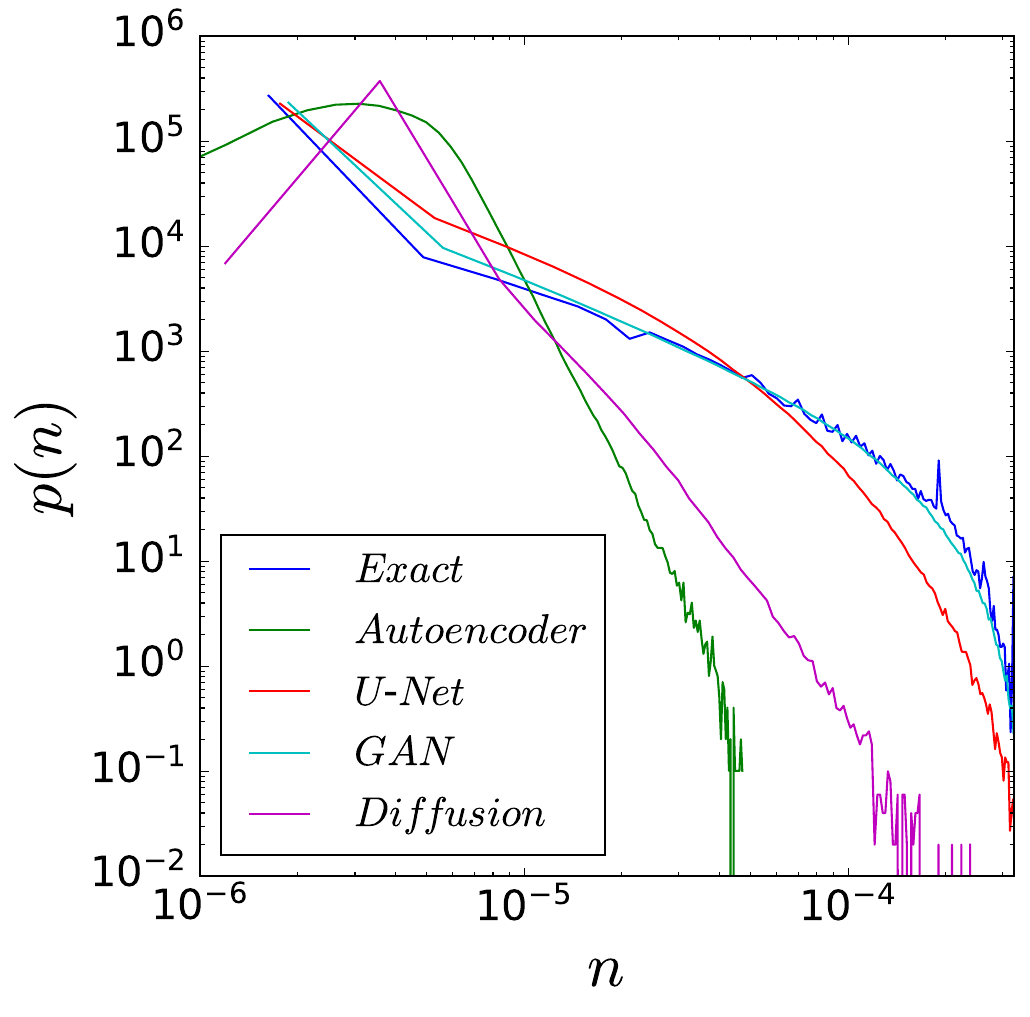}
(b)\includegraphics[width=0.45\linewidth]{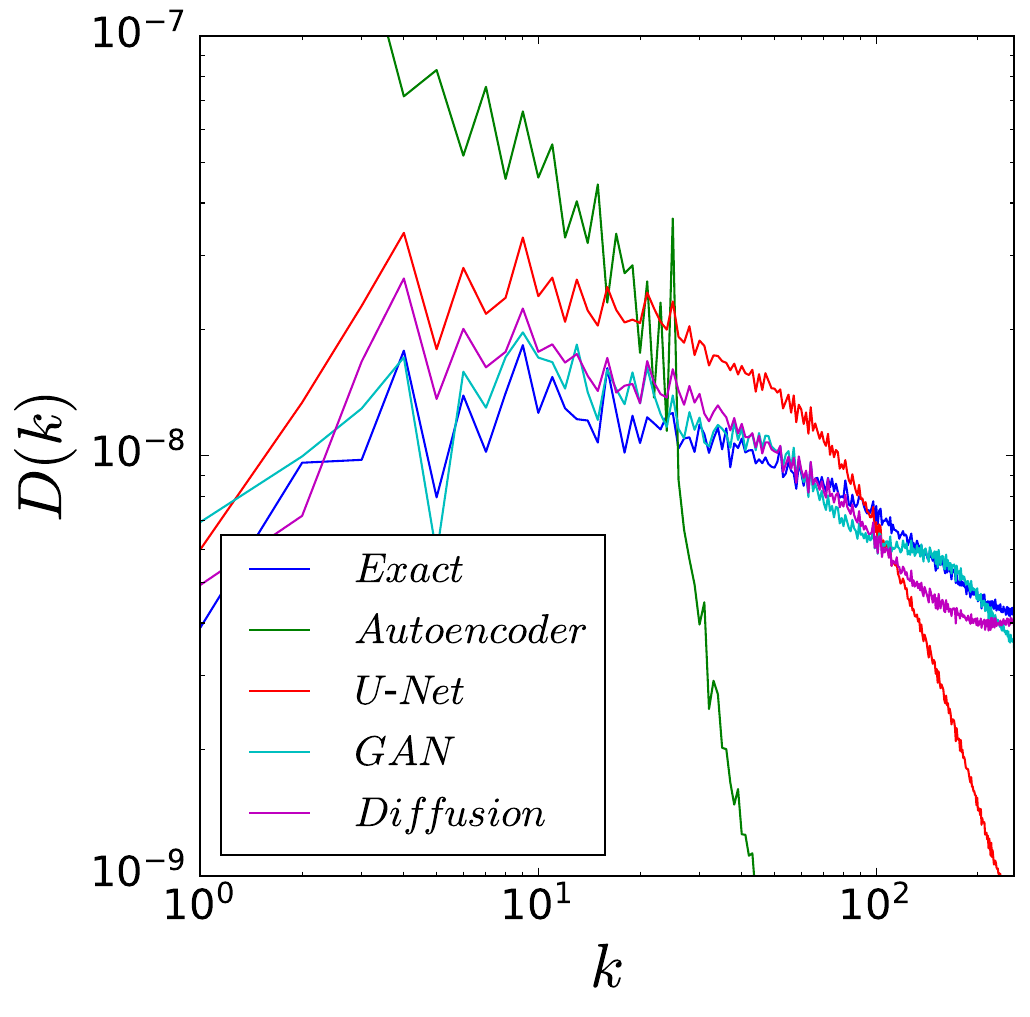}\\
(c)\includegraphics[width=0.45\linewidth]{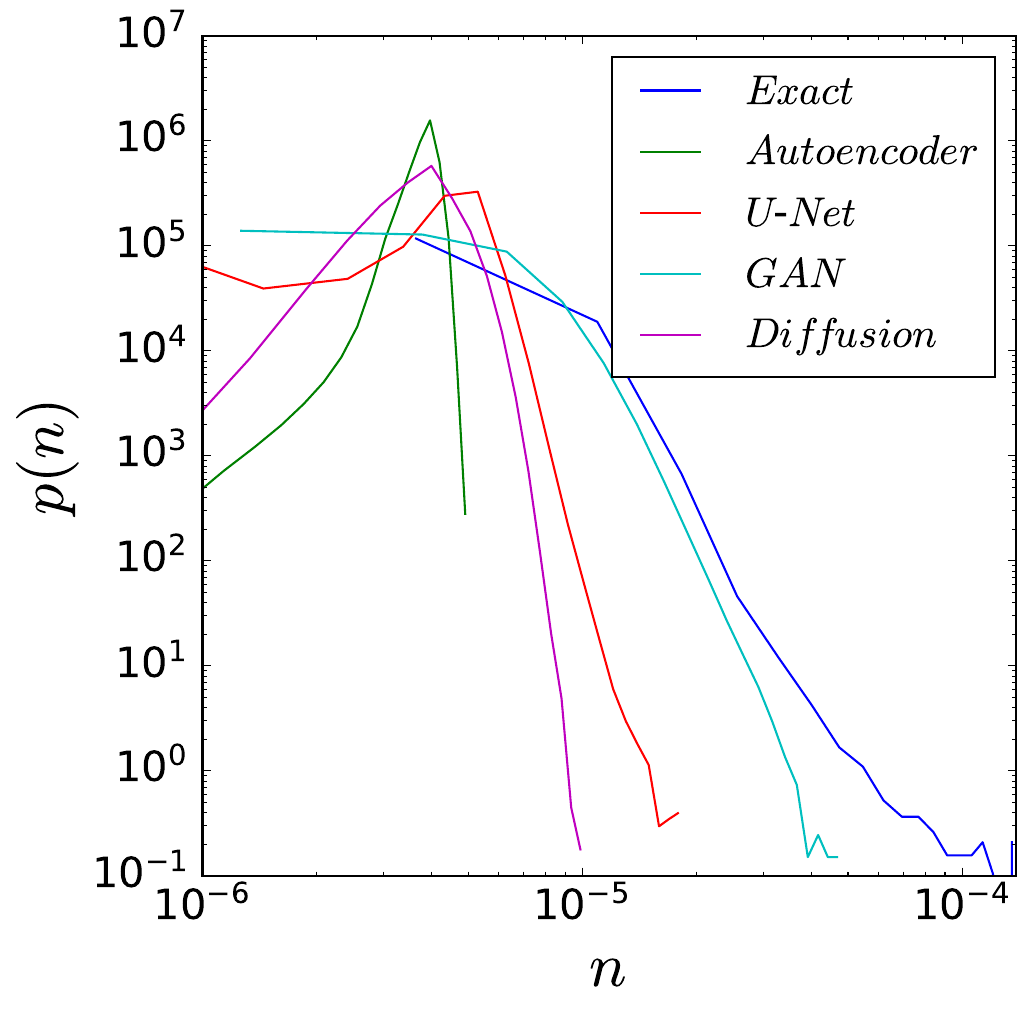}
(d)\includegraphics[width=0.45\linewidth]{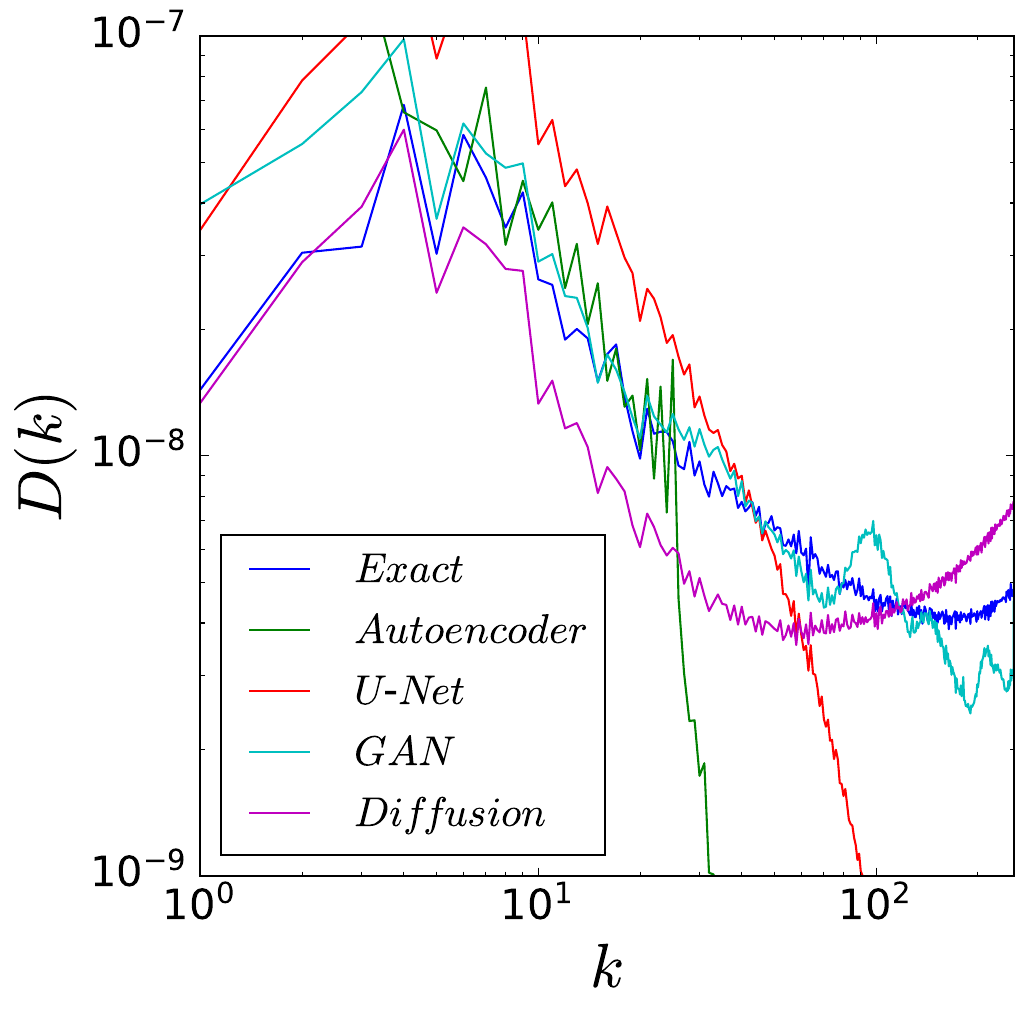}
\caption{PDFs of the (a, c) particle number density and (b, d) density spectra $D(k)$ of exact (DNS data) and predicted fields using four different architectures (Autoencoder, U-Net, GAN and Diffusion model) for (a, b) $St = 1$ and (c, d) $St = 0.05$. 
}
\label{fig:PDF_PS_St1_St005}
\end{figure}

Figure \ref{fig:PDF_PS_St1_St005} shows the PDFs of the (a, c) particle number density and (b, d) density spectra of exact (DNS data) and predicted fields using four different architectures for (a, b) $St = 1$ and (c, d) $St = 0.05$ using different network architectures averaged on 80 test data.
Results for other Stokes numbers, while not discussed in detail, can be found in the Appendix.
The density spectrum is computed by taking the modulus square of the Fourier transformed density and then summed over concentric circles in wavenumber space.

For the particle number density prediction at $St=1$, the PDF of the GAN prediction is almost perfectly superimposed with the exact DNS values. 
The U--Net model demonstrates a slight over-prediction of \thibault{small} densities, consequently under-predicting higher densities. 
Due to the background noise seen in Figure \ref{fig:visualizations_predictied_density}, the diffusion model distinctly underestimates completely void regions and overestimates regions with a low but non-zero particle number. Additionally, the PDF narrows significantly for medium and high particle densities. 
For the autoencoder, the PDF declines even more rapidly, underestimating densities. 
Similar trends can be observed for $St=0.05$.

Regarding the density spectra for GAN predictions at $St=1$, the spectrum closely approximates the exact DNS values with some oscillations.
A similar trend is observed for the diffusion model, with the exception that there is a slight under-prediction of high-frequency amplitudes. 
For the U--Net, the slope of the curve for wavenumbers less than approximately 60 is similar to those of the exact values, indicating accurate large-scale structure prediction, consistent with Figure \ref{fig:visualizations_predictied_density}. However, past wavenumber 60, there is a rapid decrease in the amplitude. 
Lastly, for the autoencoder, the spectrum deviates significantly from the exact values, with an exceedingly rapid decay at larger 
wavenumbers.
For $St=0.05$, we observe behaviors similar to those described for $St=1$, with a few exceptions. Firstly, the GAN predictions exhibit an even greater degree of oscillations. Secondly, the density spectrum for the diffusion model prediction starts to rise at higher frequencies. This unexpected rise in the spectrum is attributable to the residual noise that was not adequately eliminated during the training phase of the diffusion model. As a result, the noise becomes more predominant, skewing the signal-to-noise ratio in favor of noise and leading to this observed uptick in the density spectrum.

Theoretically, in the context of one-way coupled point-particle simulations, the particle distribution is deterministic for a given simulation, provided the number of particles is sufficiently large to consider the medium as continuous. However, the particle positions are intricately linked to the fluid's historical behavior. This introduces a bias in the particle distribution; it deviates from being uniquely dependent on regions of low vorticity and vice versa. Specifically, some particles may be entirely absent from certain low-vorticity regions due to the historical flow trajectories that have previously expelled them. Consequently, predicting the particle distribution becomes a complex problem, influenced by both current conditions and 
history effects in the flow.



\subsection{Supersampling}

As previously noted, the complexity of predicting particle distribution arises not only from current flow conditions but also from the time history of the carrier flow. 
\thibault{One approach to potentially address the computational complexity arising from a large number of particles is to explore the concept of supersampling.}
In this scenario, we consider a simulation with $10^3$ particles and aim to predict how the particle distribution would look like if there were $10^6$ particles. The primary advantage of such an approach could be the significant reduction in computational runtime by running simulations with a fewer number of particles. 
Note that this approach might be limited to one-way coupled scenarios where the back-ground flow field is common regardless of the number of particles. 
Thus, supersampling could offer a promizing compromise between computational cost and predictive accuracy, especially for high-resolution fluid-particle interaction simulations.

\begin{figure} 
\centering
(a)\includegraphics[width=0.40\linewidth]{FIGURES/vorticity_pos_neg_Exact.jpeg}
(b)\includegraphics[width=0.40\linewidth]{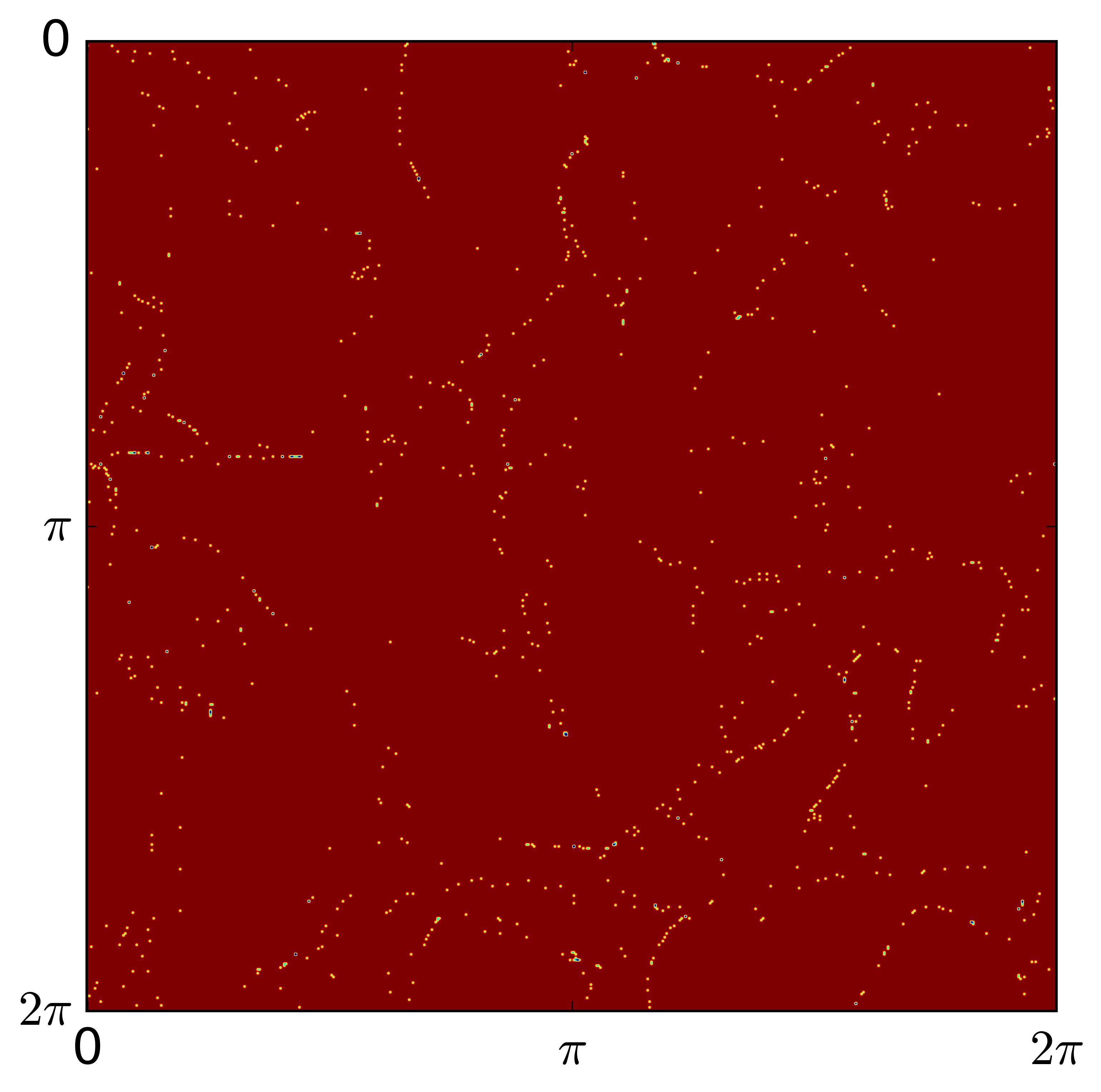}\\
\hspace{0.78cm}\includegraphics[width=0.39\linewidth]{FIGURES/colorbar_vorticity_pos_neg_Exact.jpeg}
\hspace{0.65cm}\includegraphics[width=0.39\linewidth]{FIGURES/colorbar_density_Exact.jpeg}\\
(c)\includegraphics[width=0.40\linewidth]{FIGURES/density_Exact.jpeg}
(d)\includegraphics[width=0.40\linewidth]{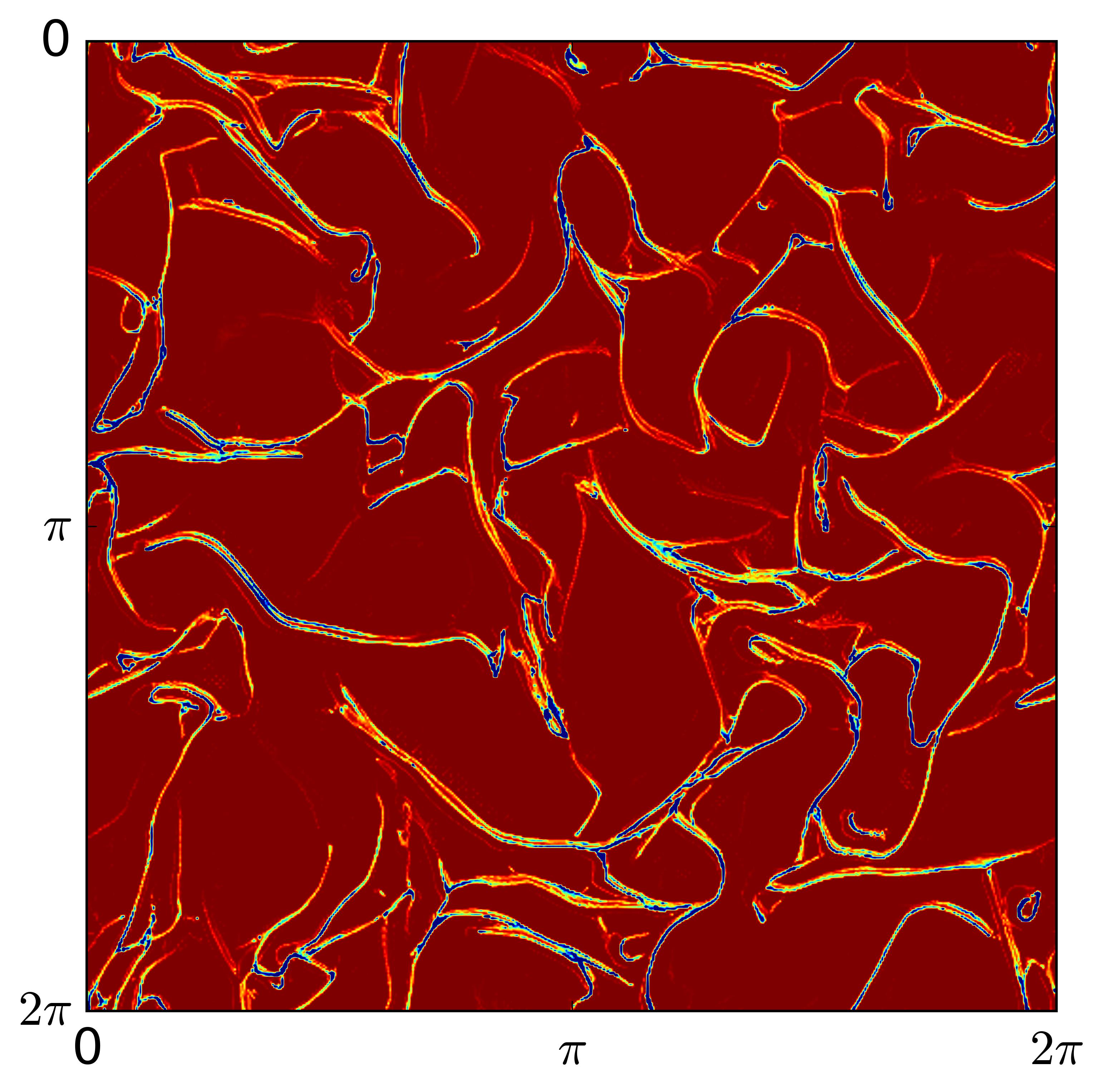}\\
(e)\includegraphics[width=0.40\linewidth]{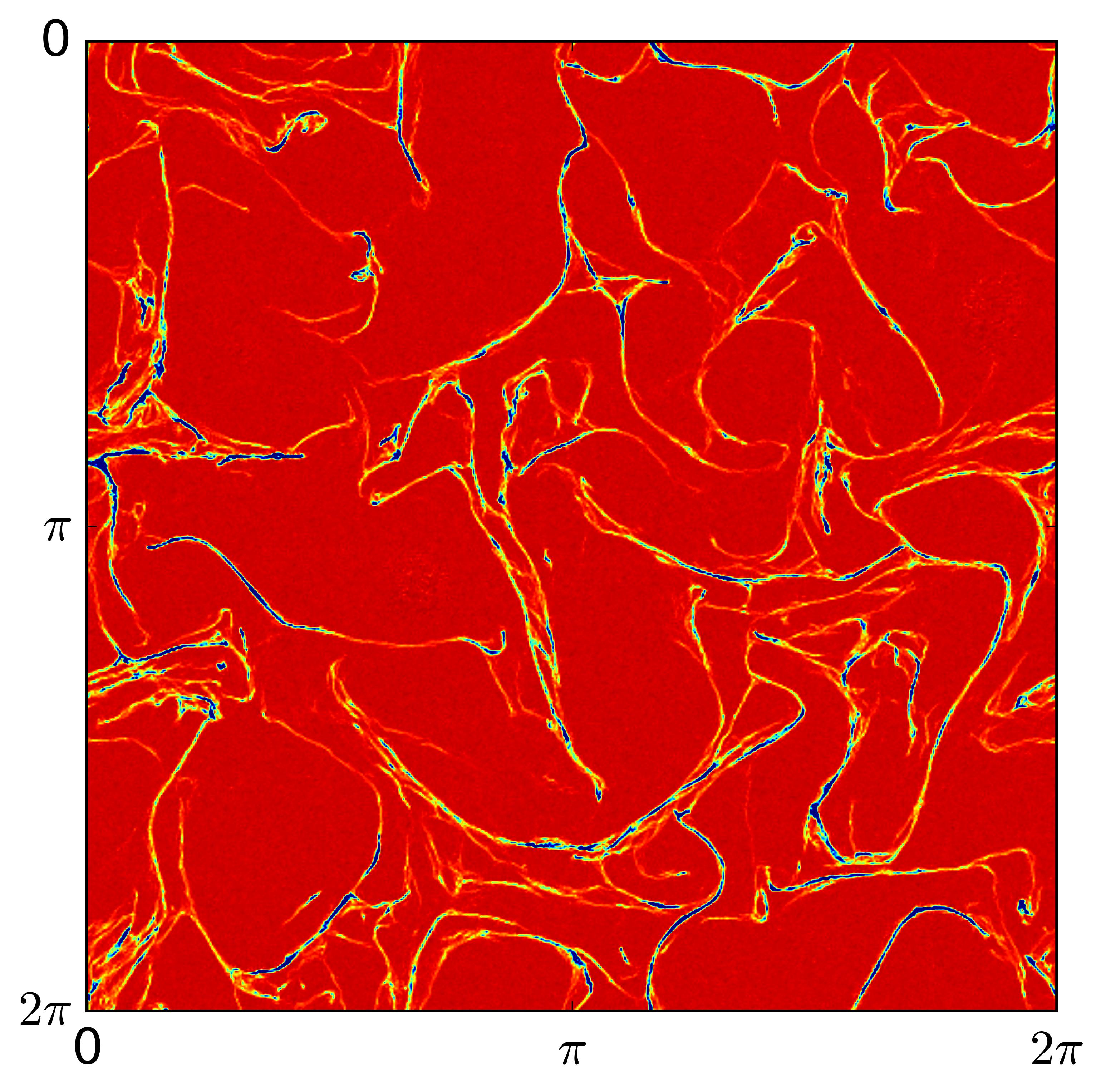}
(f)\includegraphics[width=0.40\linewidth]{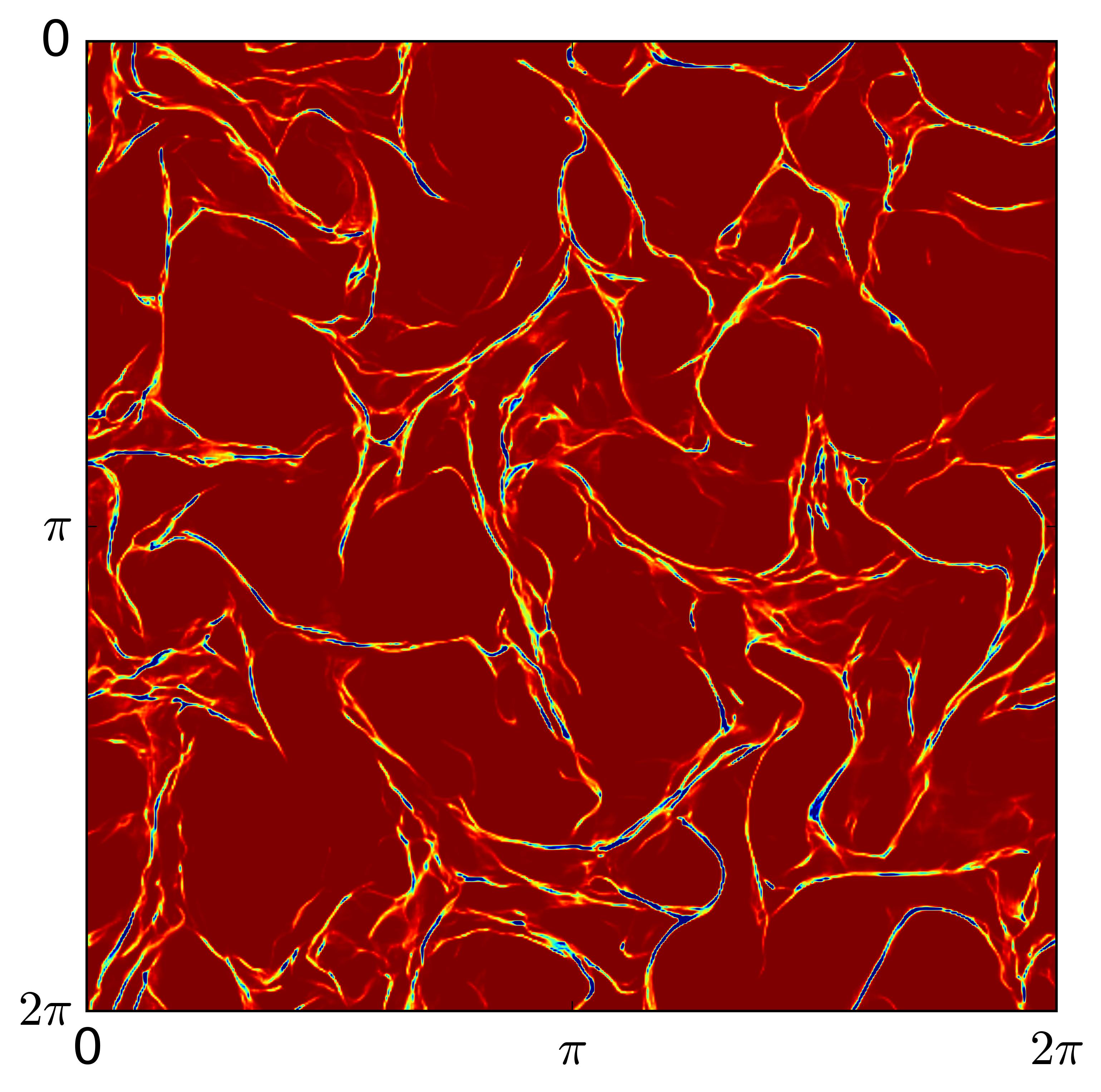}
\caption{Input (a) vorticity and (b) sub-sampled particle number density,  
ground truth (c) particle number density for $St = 1$ and predicted from vorticity using (d) GAN, (e) diffusion model and (f) U--Net. } 
\label{fig:supersampling_visu}
\end{figure}


Figure \ref{fig:supersampling_visu} shows the input (a) vorticity and (b) sub-sampled particle number density, ground truth (c) particle number density for $St = 1$ and predictions derived from vorticity using various models: (d) GAN, (e) diffusion model, and (f) U--Net. 
We observe that the particle number density distribution predicted by the GAN closely matches the exact values, even though the alignment is not identical on a pixel-by-pixel basis. Similarly, the diffusion model yields a comparable particle number density distribution; however, a persistent background noise remains evident. In the case of U--Net, we can observe finer filaments as for the original problem.

\begin{figure} 
\centering
(a)\includegraphics[width=0.45\linewidth]{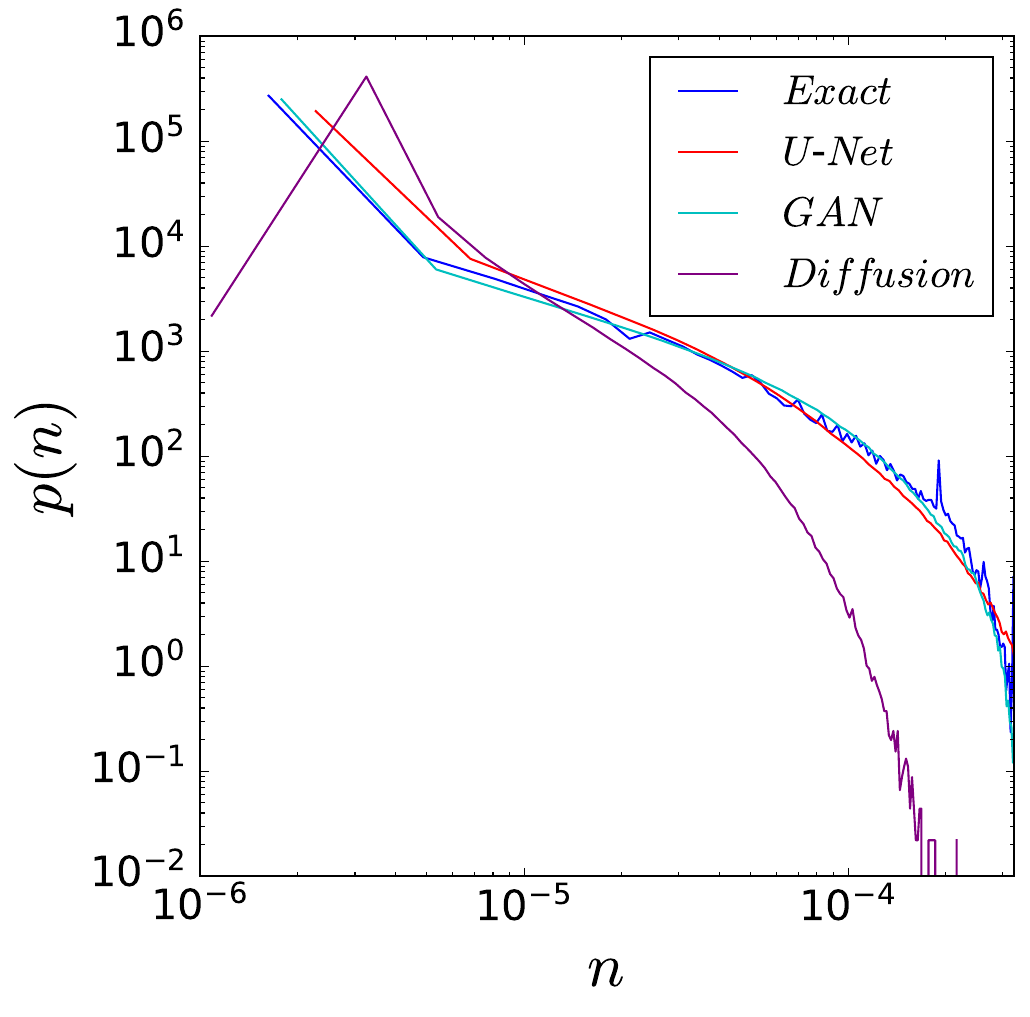}
(b)\includegraphics[width=0.45\linewidth]{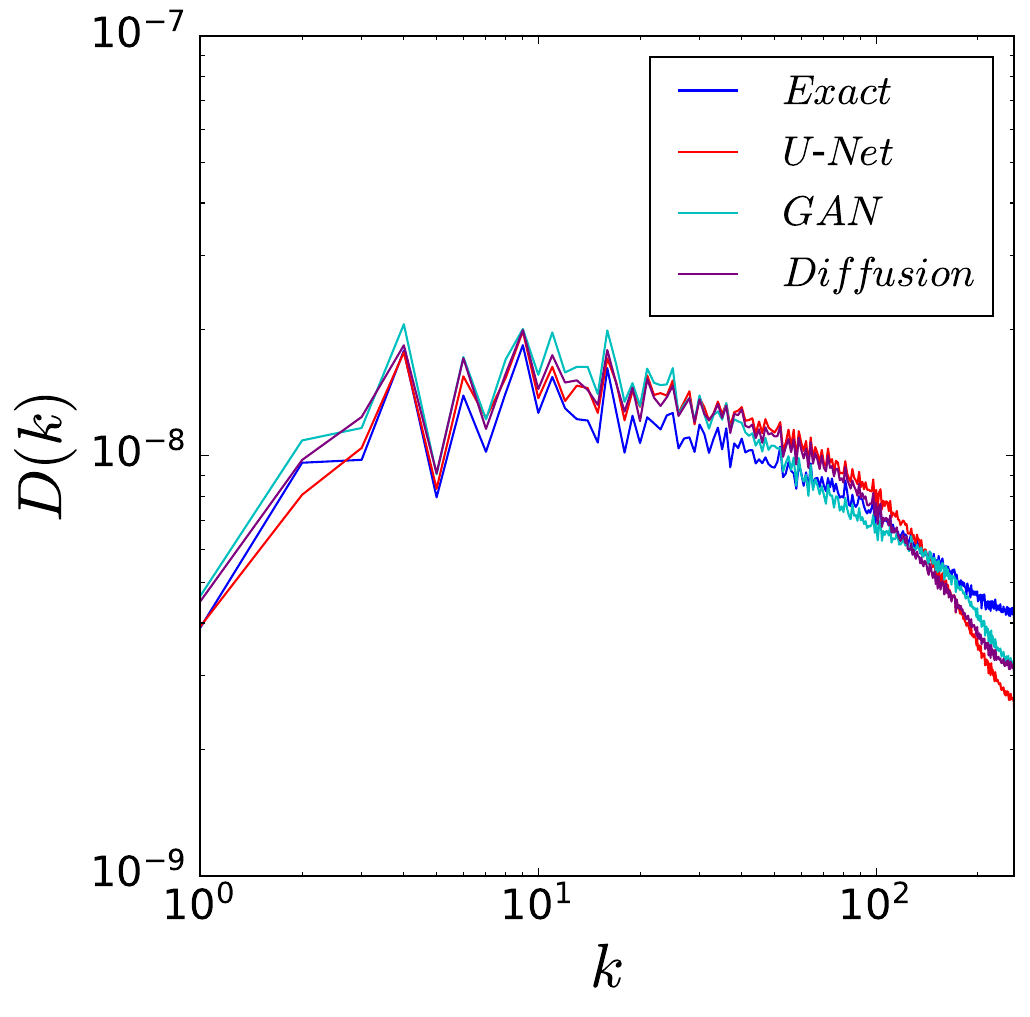}
\caption{PDFs of the (a) particle number density and (b) density spectra of exact (DNS data) and predicted fields using GAN, diffusion model and U--Net architectures for $St = 1$.} 
\label{fig:supersampling_PDF_PS}
\end{figure}

Figure \ref{fig:supersampling_PDF_PS} shows PDFs of the (a) particle number density and (b) density spectra of DNS data and predicted fields using three different architectures for $St = 1$.
For the GAN, the PDF continues to be nearly perfectly superimposed on the exact DNS values. 
A similar trend is observed for U--Net, where the PDF now is also superimposed almost with the exact data, representing an improvement in capturing both average and high-density regions.
The diffusion model shows marked improvements, especially in the prediction of high particle densities, although it still falls short in accurately capturing entirely void regions, presumably due to persistent background noise.
Regarding the density spectra, no substantial variations are noted for both the GAN and the diffusion model.
However, for U--Net, the predicted spectrum has moved closer to the exact DNS values, indicating improved prediction accuracy across different scales.

To summarize, we find that the predictive capability of the neural networks improve significantly when the information about a few particles are available and can be used as the input for the networks along with the vorticity data, as opposed to using only the vorticity data as the input. Hence these networks can be used as a post-processing tool to extract higher-resolution particle data by performing numerical simulations with only a few particles, which will reduce the cost of the simulations. 
\thibault{Further investigations are necessary concerning the dependence of the input number of particles on the prediction.}

\subsection{Vorticity prediction}

We are also interested in knowing the feasibility of the use of the neural networks for the enstrophy prediction from the particle number density, which could be of interest to experimentalists. 
To this end, we try to predict the enstrophy field using the particle number density fields for all five Stokes numbers available to us. 
Note that using particle number density fields from different Stokes numbers to predict a single enstrophy field is possible only in the case of one-way coupled point-particle simulations.

\begin{figure} 
\centering
(a)\includegraphics[width=0.40\linewidth]{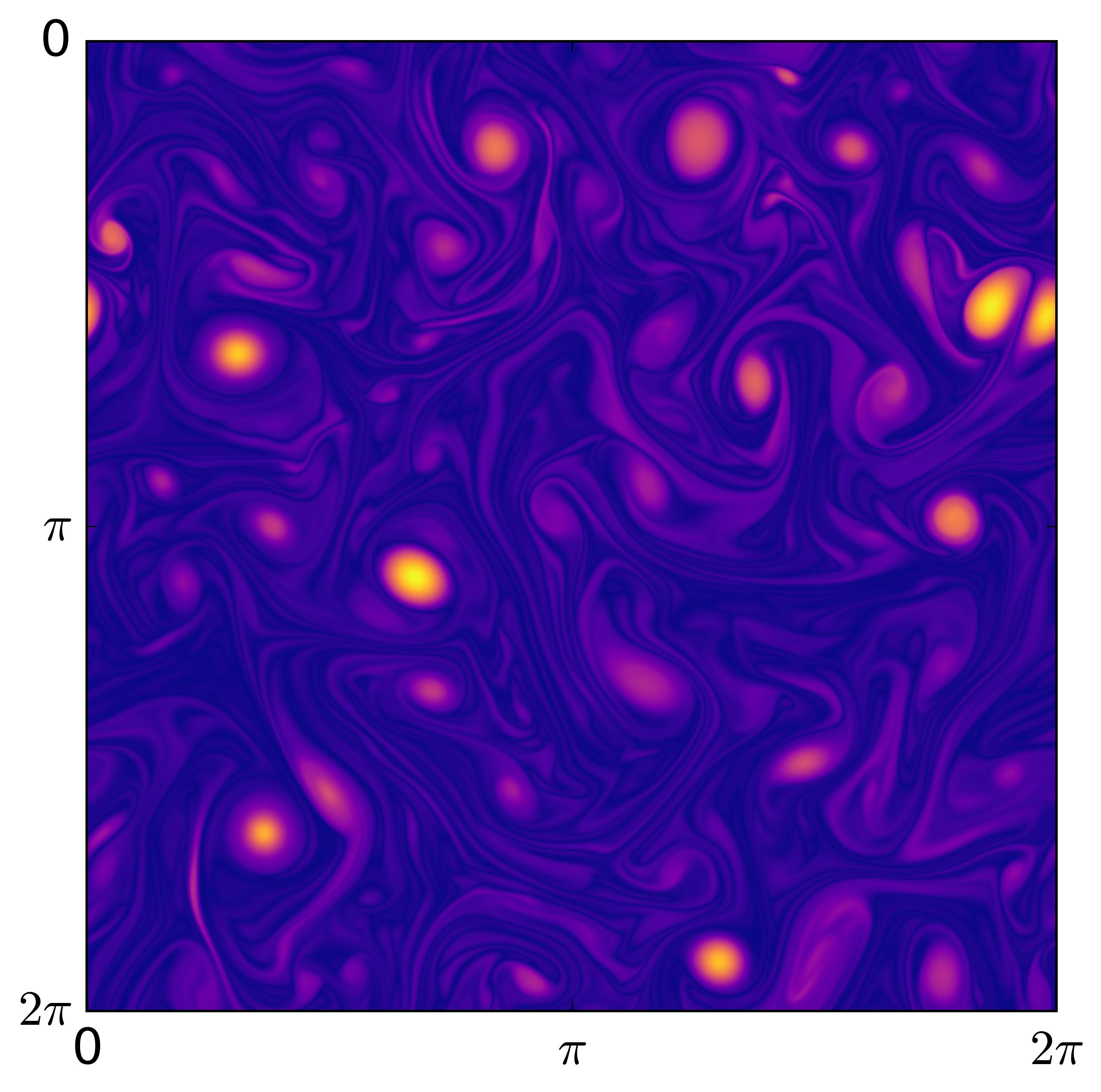}
(b)\includegraphics[width=0.40\linewidth]{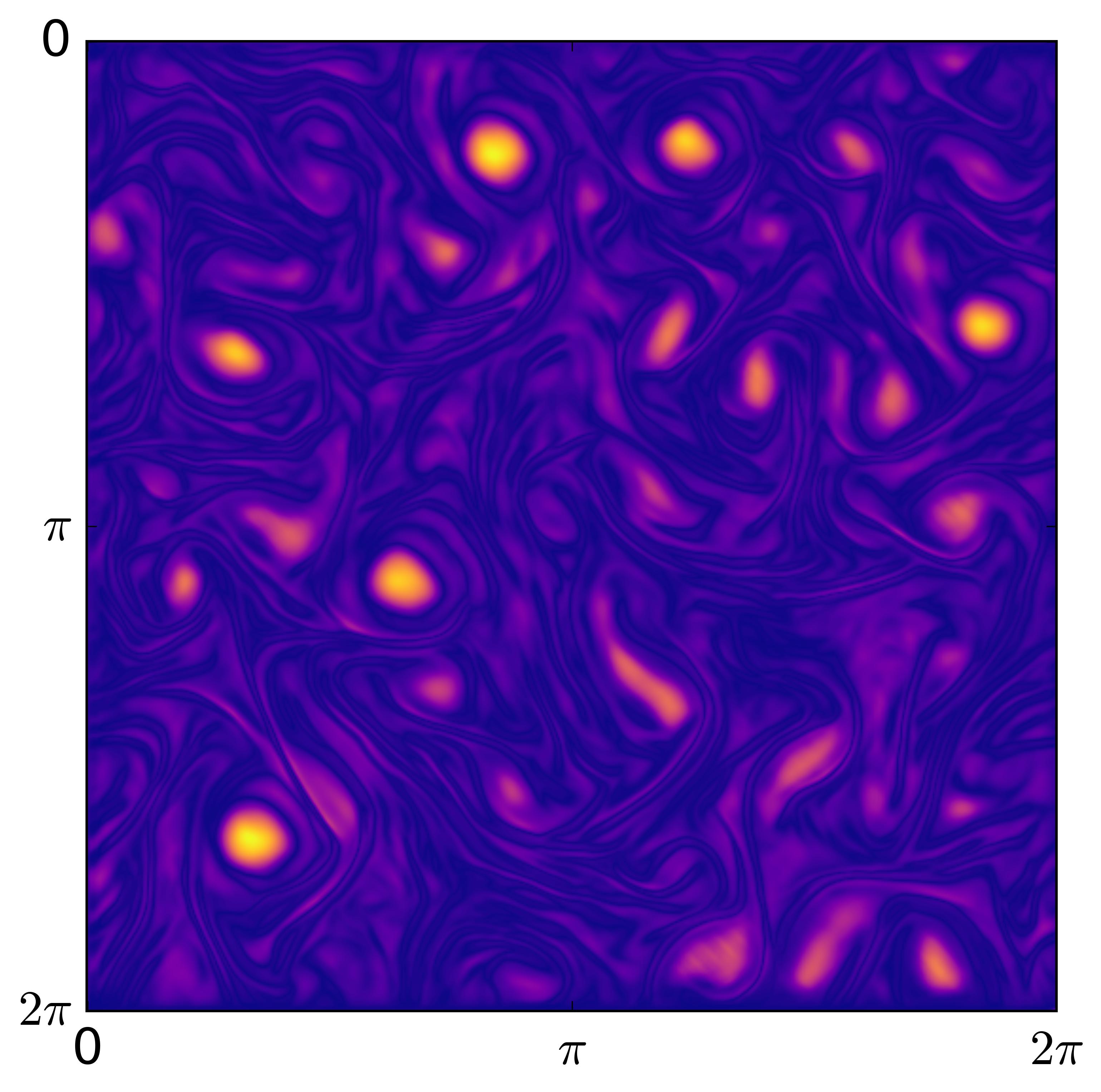}\\
(c)\includegraphics[width=0.40\linewidth]{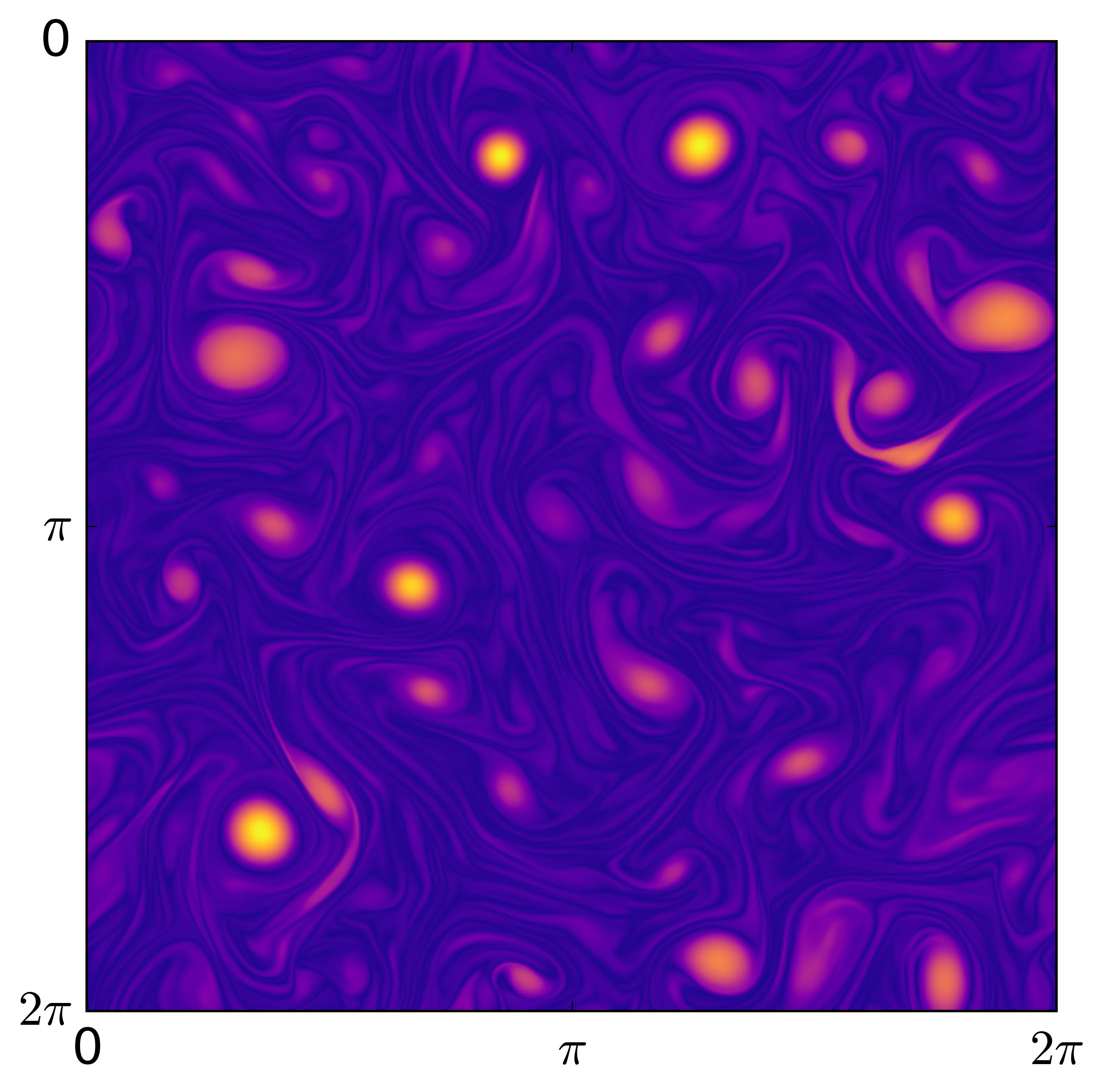}
(d)\includegraphics[width=0.40\linewidth]{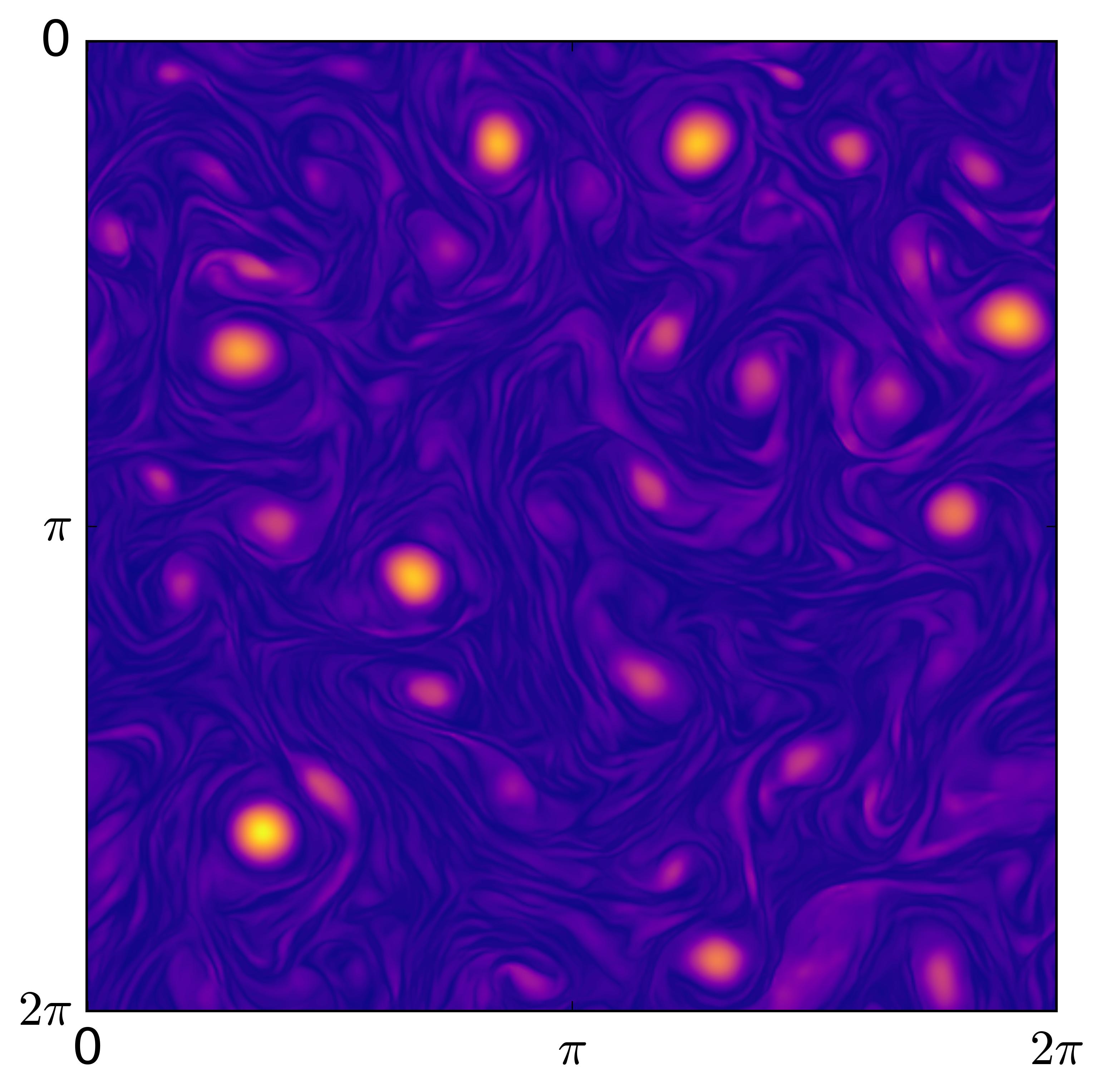}\\
\hspace{0.78cm}\includegraphics[width=0.87\linewidth]{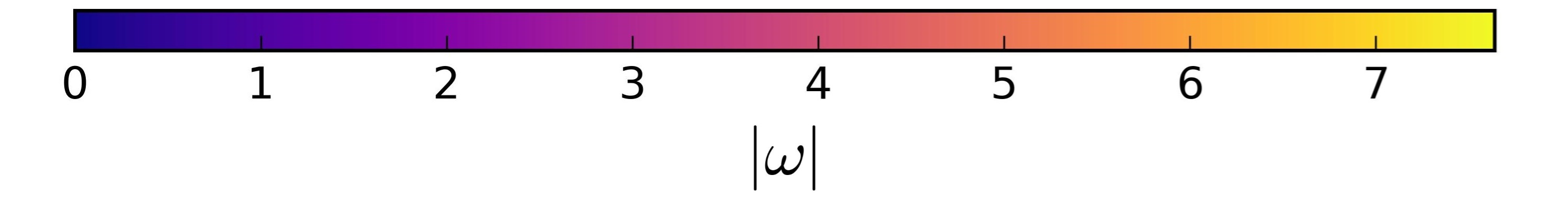}
\caption{Ground truth of (a) the absolute value of the vorticity and predicted from particle number density using (b) GAN, (c) diffusion model and (d) U--Net.} 
\label{fig:vorticity_visu}
\end{figure}

Figure \ref{fig:vorticity_visu} shows the absolute values of the vorticity for (a) the exact field, and predictions from particle number density using (b) the Generative Adversarial Network (GAN), (c) the diffusion model, and (d) the U--Net architecture. 
We can observe that the overall structure is similar to the exact field, capturing vortex locations with accurate shape, size, and magnitude. However, in the case where two vortices are in close proximity to each other, the models tend to predict a single, larger vortex instead  of the two individual structures. This behavior can be attributed to the centrifugal force that ejects particles, effectively removing data from that location and hence inhibiting the models' ability to reconstruct the absolute values of the vorticity correctly.

\begin{figure} 
\centering
\includegraphics[width=0.45\linewidth]{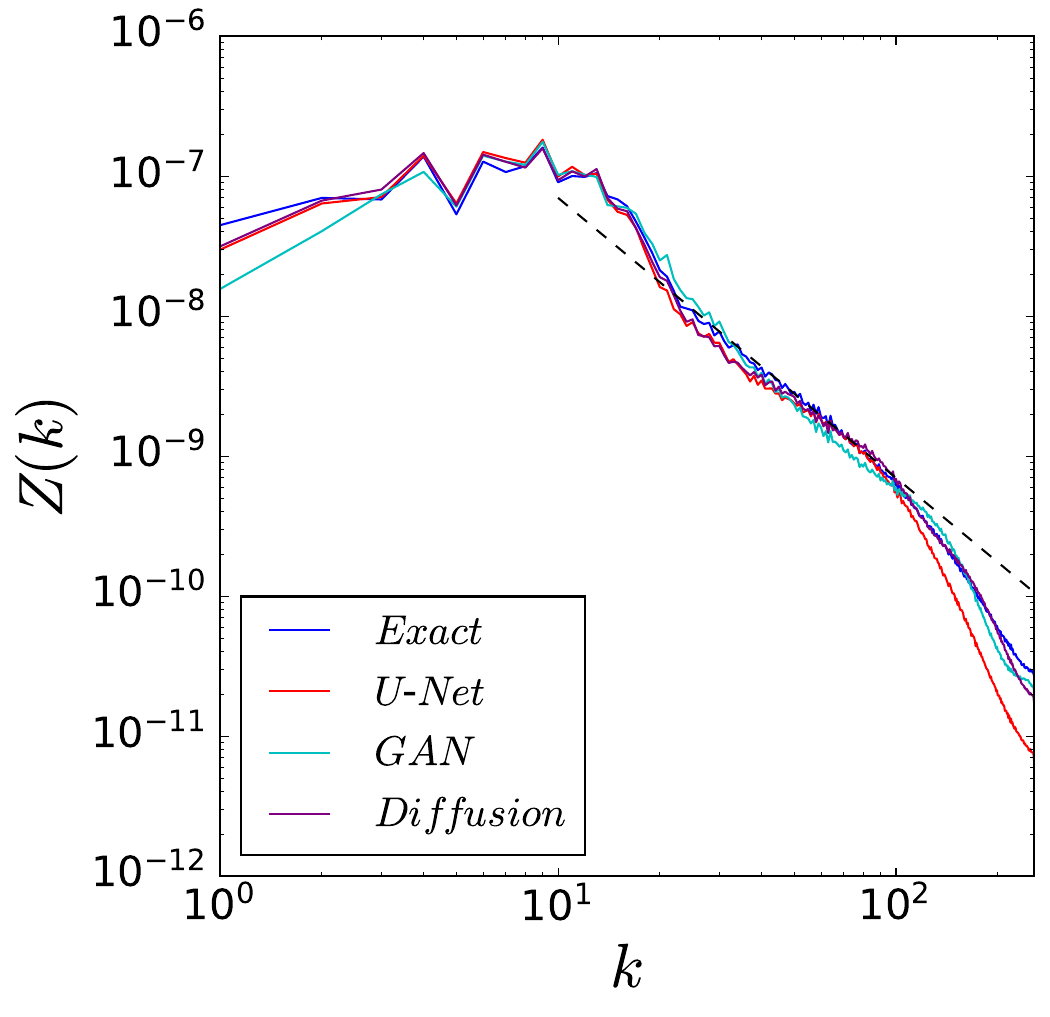}
\caption{
Enstrophy
spectra $Z(k)$ of exact (DNS data) and predicted absolute value of the vorticity fields using different architectures. Dashed line indicates the power law $k^{-2}$. 
}
\label{fig:vorticity_PS}
\end{figure}

Figure \ref{fig:vorticity_PS} shows the 
enstrophy
spectra of the exact vorticity field (DNS data) and the predicted absolute values of the vorticity using various network architectures. For both the GAN and the diffusion model, the spectra of the predicted data are similar to those of the exact values, with a few deviations. 
In contrast, the U--Net model exhibits a close approximation to the exact density spectra except at higher wavenumbers, where the predicted spectrum decreases more rapidly than it should, indicating a deviation from the exact field.




\section{Conclusions}
\label{sec:conclusions}

In this work, four neural networks 
for synthesizing preferential concentration fields of inertial particles in fully developed two-dimensional turbulence have been developed and assessed.
Instantaneous snapshots of vorticity fields of DNS data were taken as input, and particle concentration fields were generated using autoencoder, U--Net, GANs, and diffusion model. 
The quality of prediction is quantified by comparing PDFs and density spectra of the synthesized fields with the DNS data.
The best results were obtained with the GAN, showing similar cluster and void regions as present in the DNS data. This can be explained by the ability of GANs to generate fine-scale features, as well as by its generative properties.

Furthermore, the technique of supersampling has been explored 
to enhance computational efficiency. 
This allows machine learning models to converge more easily, providing them with information about the time history of the particles. 
By predicting particle distributions in simulations with fewer particles and extrapolating to higher particle counts, this approach is promising in reducing computational demands while maintaining a balance between accuracy and resource utilization. 
This also demonstrates the potential of the use of neural networks as a subgrid model in the context of large-eddy simulations where fewer particles, as opposed to full set of particles, are used in the numerical simulations.

Inverting the input and output in the proposed procedure, we showed that reconstructing the enstrophy from the particle positions (e.g., obtained via PIV data) is likewise possible. This is a physically relevant task and yields a promising application for experimentalists. 
In the future, we also plan to predict fundamental quantity quantities, such as Reynolds and Stokes numbers. 

\medskip


\section*{Acknowledgments}

T.M.O. and K.S. acknowledge funding from the Agence Nationale de la Recherche (ANR), grant ANR-20-CE46-0010-01. 
K. Matsuda acknowledges financial support from JSPS KAKENHI Grant Number JP20K04298.
S. S. Jain acknowledges funding from the Boeing Co. and support from Georgia Tech.
K. Maeda acknowledges funding from SRB Co., Ltd. and support from Purdue University.
Benjamin Kadoch and Zetao Lin are acknowledged for providing the DNS data, Sixin Zhang for fruitful discussion and valuable advice, and Katsunori Yoshimatsu for his guidance and for facilitating a portion of the computations at the Computing Center of University of Nagoya.
Centre de Calcul Intensif d’Aix-Marseille is acknowledged for granting access to its high performance computing resources.

\medskip


\bibliographystyle{abbrvnat}

\newpage

\section{Appendix}

\begin{figure} 
\centering
(a)\includegraphics[width=0.35\linewidth]{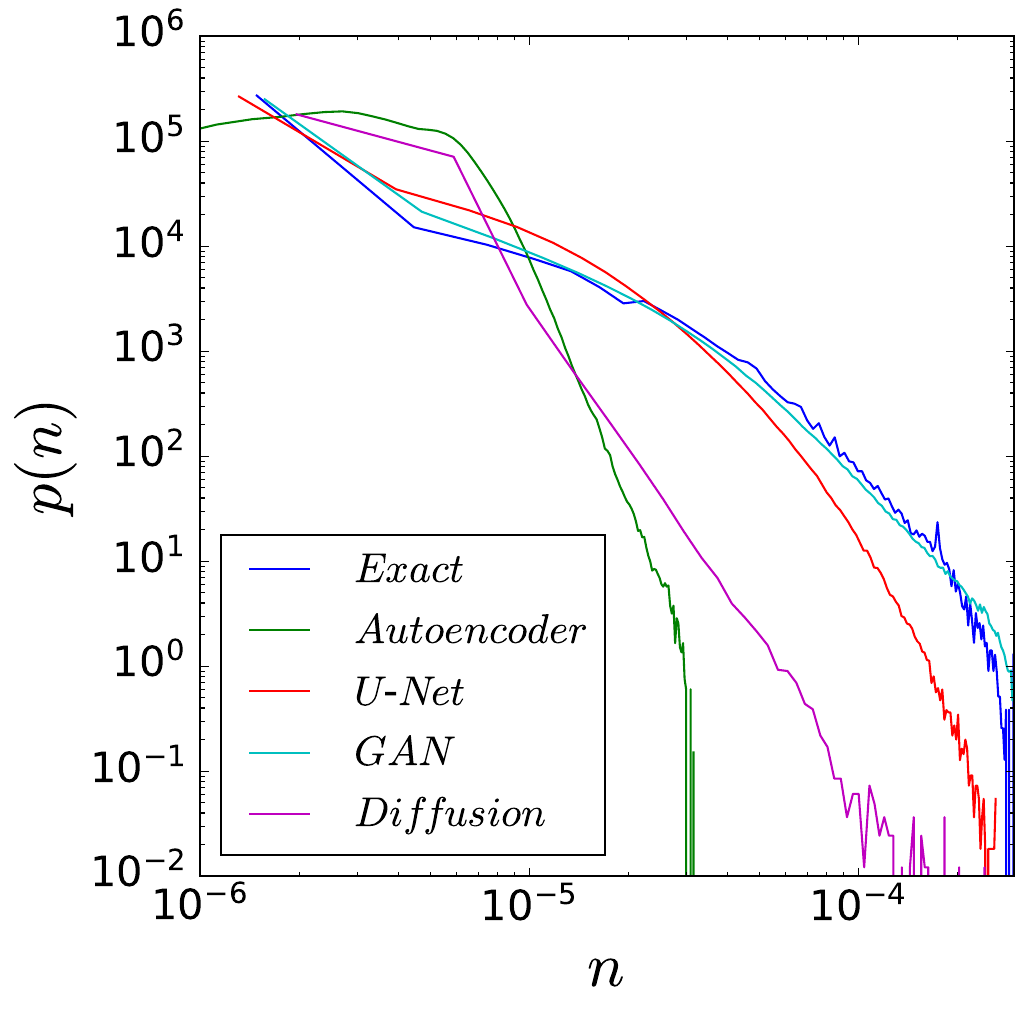}
(b)\includegraphics[width=0.35\linewidth]{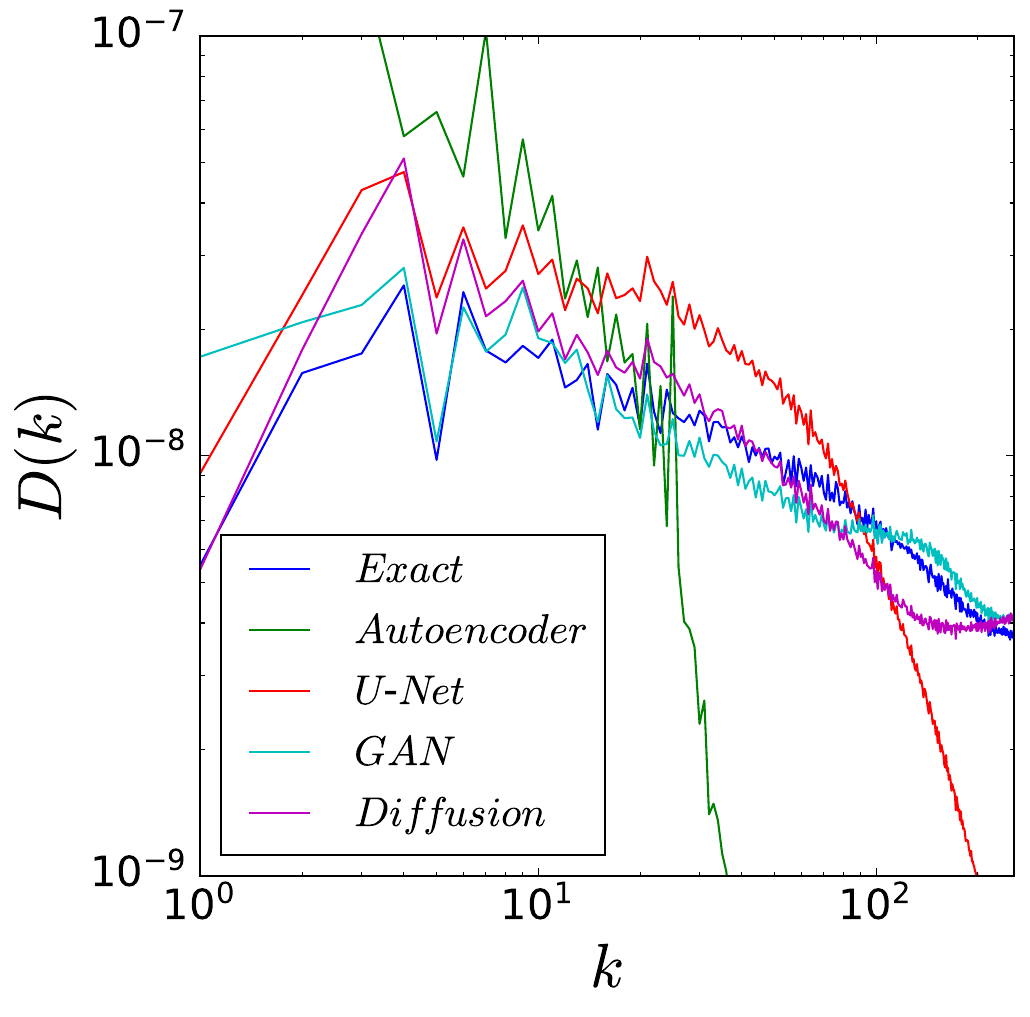}\\
(c)\includegraphics[width=0.35\linewidth]{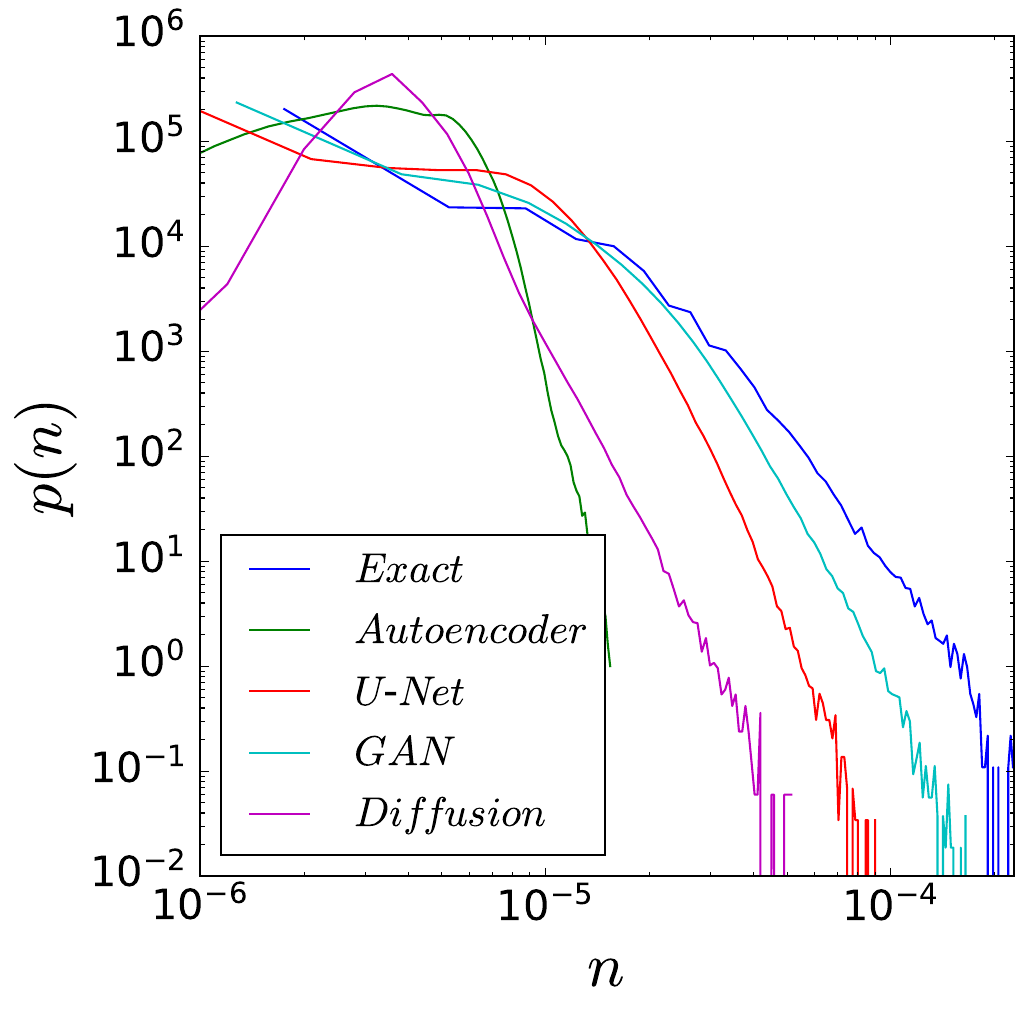}
(d)\includegraphics[width=0.35\linewidth]{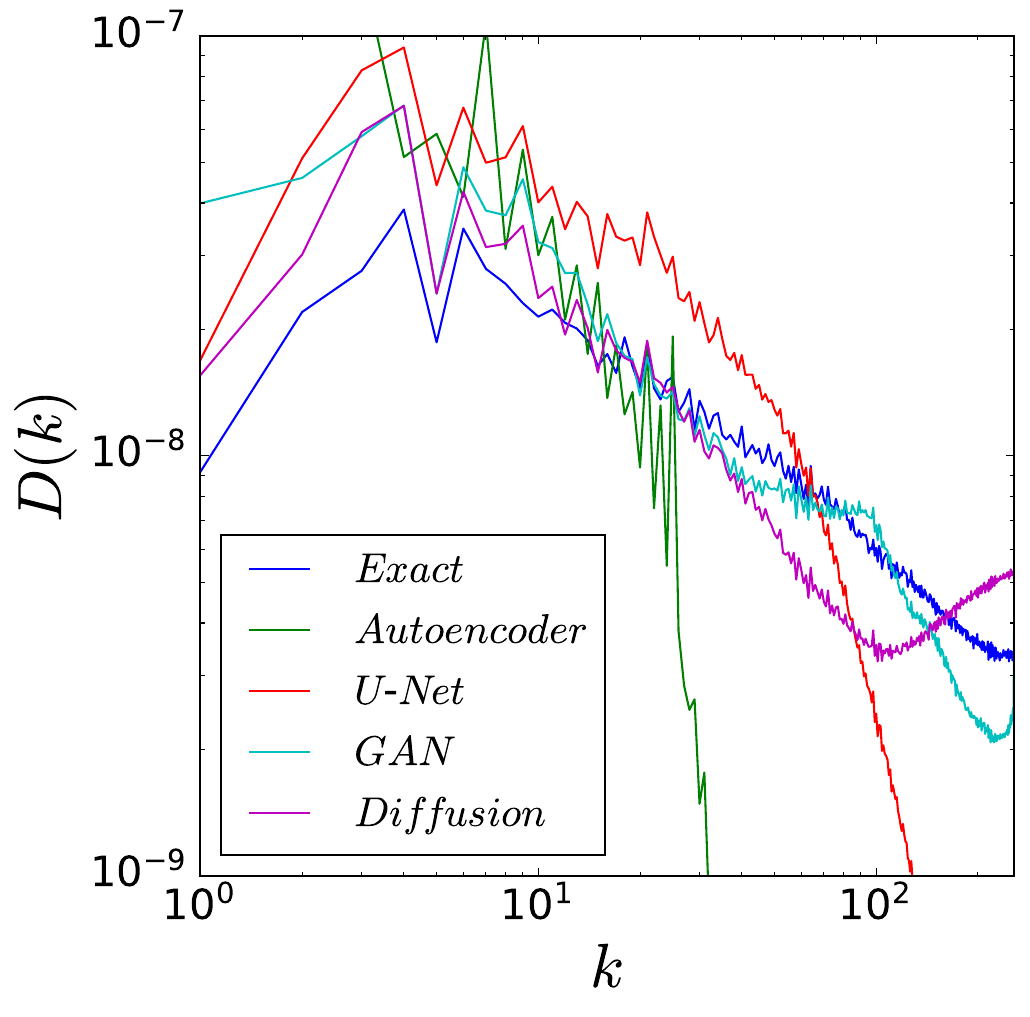}\\
(e)\includegraphics[width=0.35\linewidth]{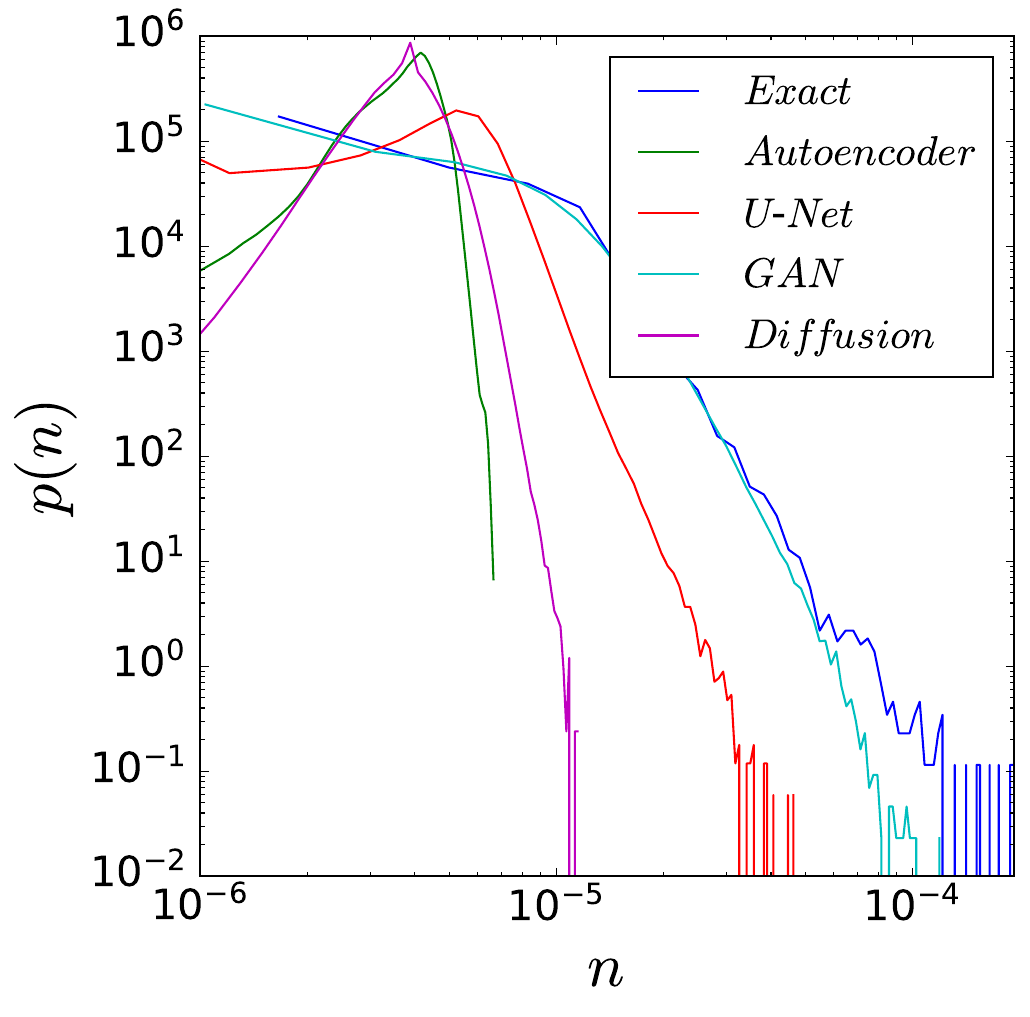}
(f)\includegraphics[width=0.35\linewidth]{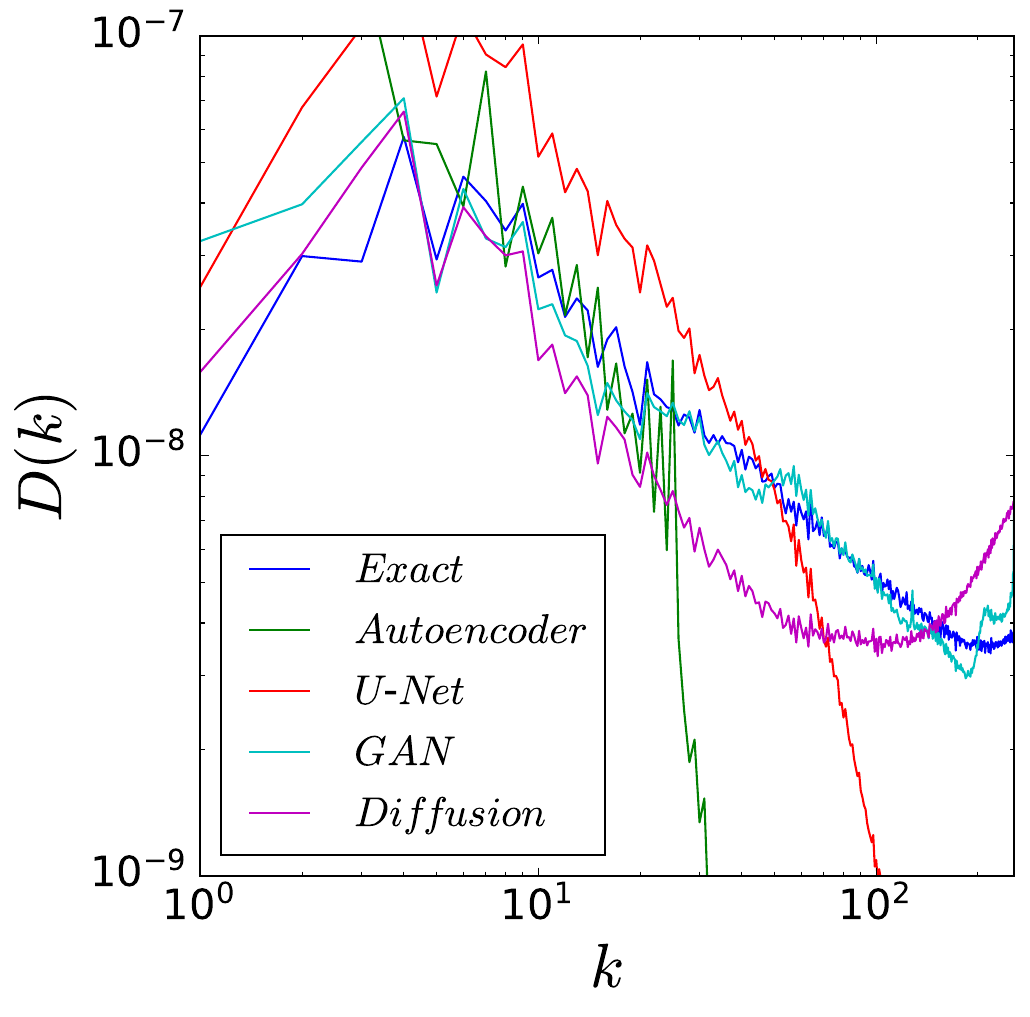}\\
\caption{PDFs of the (a, c, e) particle number density and (b, d, f) density spectra $D(k)$ of exact (DNS data) and predicted fields using four different architectures (Autoencoder, U-Net, GAN and Diffusion model) for (a, b) $St = 0.5$, (c, d) $St = 0.25$ and (e, f) $St = 0.1$.
} 
\label{fig:appendix}
\end{figure}

Here, we present different results for other Stokes numbers, i.e., $St = 0.5, 0.25$ and 0.1.
The results are similar and continuously approach the two extremes, i.e. $St = 1$ and 0.05, which are discussed in detail in the main text.
Figure~\ref{fig:appendix} shows corresponding PDFs of particle number density and density spectra for the different architectures in comparison with the DNS results.

\end{document}